\documentclass[prd,
showpacs,preprintnumbers,nofootinbib]{revtex4-1}

\usepackage{latexsym}
\usepackage{amssymb}
\usepackage{amsmath}
\usepackage[dvips]{epsfig}
\usepackage{color}
\usepackage{ulem}
\usepackage{mathtools}

\usepackage{array,booktabs}
\newcolumntype{C}{>{$}c<{$}}
\AtBeginDocument{
\heavyrulewidth=.08em
\lightrulewidth=.05em
\cmidrulewidth=.03em
\belowrulesep=.65ex
\belowbottomsep=0pt
\aboverulesep=.4ex
\abovetopsep=0pt
\cmidrulesep=\doublerulesep
\cmidrulekern=.5em
\defaultaddspace=.5em
}

\newcommand{\il}{\raisebox{1.7pt}{$\scriptstyle\blacktriangleright$}\;}

\usepackage{epstopdf}
\usepackage{colortbl}
\makeatletter

    \def\CT@@do@color{%
      \global\let\CT@do@color\relax
            \@tempdima\wd\z@
            \advance\@tempdima\@tempdimb
            \advance\@tempdima\@tempdimc
    \advance\@tempdimb\tabcolsep
    \advance\@tempdimc\tabcolsep
    \advance\@tempdima2\tabcolsep
            \kern-\@tempdimb
            \leaders\vrule
                    \hskip\@tempdima\@plus  1fill
            \kern-\@tempdimc
            \hskip-\wd\z@ \@plus -1fill }
    \makeatother

\usepackage{multirow}

\renewcommand{\mathbf}{\boldsymbol}

\newcommand{\phisq}{\big \langle \phi^{\kn 2} \big\rangle_{\rm ren}}

\newcommand{\bs}{\beta^{{\kern 0.5pt}\prime}}

\newcommand{\ds}{\displaystyle}

\newcommand{\kn}{{\kern 1pt}}
\newcommand{\akn}{{\kern -1pt}}

\newcommand{\e}{\mathrm{e}}

\newcommand{\pp}{\, ... \,}
\newcommand{\nn}{\nonumber}

\newcommand{\ren}{{\rm ren}}

\newcommand{\tcs}{\tabcolsep}
\newcommand{\acs}{\arraycolsep}

\newcommand{\erh}{\extrarowheight}

 \numberwithin{equation}{section}

 \definecolor{superlight-gray}{gray}{0.9}

\begin {document}
\title{Vacuum polarization effects of pointlike impurity}
\author{Y.V.Grats\footnote{E-mail: grats@phys.msu.ru},
P.\,Spirin\footnote{E-mail: pspirin@physics.uoc.gr}} \affiliation{
Department of Theoretical Physics, M.V.Lomonosov Moscow State University, 119991,
Moscow, Russian Federation} 


\begin{abstract}
 We develop precise formulation for the effects of vacuum polarization  near a pointlike source with a zero-range ($\delta$-like) potential in three spatial dimensions.
 There are different ways of introducing
$\delta$-interaction in the framework of  quantum theory. We
discuss the approach based on the concept of self-adjoint extensions of  densely defined symmetric operators.
Within this approach we consider the
real massive scalar field in three-dimensional Euclidean space with
a single extracted point. Appropriate boundary conditions imposed at this point enable one to consider all self-adjoint extensions of $- \Delta$ as operators which can describe a pointlike source with a zero-range potential.
In this framework we compute
the renormalized vacuum expectation value of the field square
$\langle \phi^{\kn 2}(x)\rangle_{\rm ren}$ and the renormalized
vacuum average of the scalar-field's energy-momentum tensor
$\langle T_{\mu\nu}(x)\rangle_{\rm ren}$. Asymptotic cases are discussed in detail.
\\[7pt]
\end{abstract}
\maketitle


 \section{Introduction}

Seventy five years ago it was predicted some phenomenon, named ''the Casimir effect'' later in honour of the author.
Being confirmed experimentally ten years later,   it became a direct evidence of a close connection between quantum effects and macroscopic external conditions. These ones may be manifested, e.g. as nontrivial local/global  structure  of the background manifold, as boundaries or external fields, etc. Subsequently, a number of new results was obtained in the study of processes on various spatio-temporal scales, at a level of both experimental and theoretical studies. And now the Casimir effect is a subject of extensive research  by
specialists in the quantum field theory or physics of condensed matter, by scientists
working in various fields of gravity and cosmology.

The Casimir effect is certainly one of the most striking manifestations of vacuum polarization, which is the standard problem of research within the quantum
theory. These studies revealed a non-trivial dependence of vacuum effects on external conditions.
Such a relation is explained by the fact that solutions to the field equations essentially depend upon  the global structure of the background  manifold. This effective non-locality leads to some non-trivial effects even when one considers the  Euclidean space with a
discrete number (finite or infinite) of extracted points. In this
case,   adequate boundary conditions enable to consider quantum
systems given by a Hamiltonian, which is heuristically given by the expression
\begin{align}\label{eq1jj}
H=-\Delta+\sum\limits_{y}\,\lambda_y\,
\delta^d(\boldsymbol{x}-\boldsymbol{y})\, ,
\end{align}
where $\Delta$ denotes the
self-adjoint Laplacian in $L^2({\mathbb{R}}^d)$  with definition domain $H^{2,2}({\mathbb{R}}^d)$, $d=1, 2, 3$  is
  dimension of the configuration space,
$\delta^d(\boldsymbol{x}-\boldsymbol{y})$ --- the Dirac's $\delta$-function
localized at $\boldsymbol{y}$, and $\lambda_y$ stands for corresponding coupling
constant.

Models of this type are known in the literature as ''contact
interaction models'', ''zero-range interaction models'',
''delta-like interaction models'', ''point interaction models''
and ''Fermi potential models''. These models  have been extensively
discussed in physical and mathematical literature in connection
with Kronig\,--\,Penney model \cite{Kronig}, in description of
short-range nuclear forces \cite{Demkov}, etc.

 Encoding of the
boundary conditions by a $\delta$-like potential made it possible
to investigate the Casimir interaction of two parallel perfectly
conducting plates \cite{Mamaev,Mamaev2,Bordag92,Bordag,BordagBook,Castaneda1,Grats2018}, two point or
linear sources with a zero-range potential \cite{Grats2019,Fermi1,Fermi2,Fermi3}, and so on.

Similar situations take place
near conical topological defects. Indeed, the transverse cross section of  the cosmic string space-time has a conical structure with a
$\delta$-like singularity at the vertex of a cone \cite{Sokolov77,Vilenkin2},
and in conformal  coordinates the equation for a massless scalar field
with non-minimal  coupling takes the form of eq.\,(\ref{eq1jj}) \cite{Grats17,Grats20}.
Notice, that in consideration of field processes near cosmic-strings, as well as on the analogous (2+1)-dimensional spaces \cite{Kay, Jackiw, Jackiw-rev, Allen90}, it was crucial a question, in what sense  one should regard the expression (\ref{eq1jj}). In particular, by one way or another, one can propose to smooth out the singularity and to regard the
$\lambda\kn\delta$-potential as a distributional limit of an appropriate delta-like sequence \cite{Allen90,Allen96,Kay,Khus04,Bezz06,Bezz15}. Some conceptual problems of  the $\lambda\kn\delta$-potential introduction were discussed in \cite{Berezin,Zeldo}.

 In recent literature, various singular configurations have been studied, including point-like interactions \cite{Solod}, delta potentials at finite temperature \cite{Spreafico}, Casimir interactions of manifolds with different codimensions \cite{Scardicchio}, Casimir effect with a boundary hyperplane \cite{Albus2010,Bezz23}, and Casimir problems approached using Zeta-regularization \cite{Fermi2,Fermi3}.

 We
follow the approach based on the concept of self-adjoint extensions of a densely defined  symmetric operator.

It is known that in dimensions $d\geqslant 4$, any self-adjoint extension of $H$  simplifies to $-\Delta$ itself. In dimensions $d=2,3$, the free operator $-\Delta$ is self-adjoint when acting on finite functions only. In these dimensions, the self-adjointness of $H$ may be preserved even when corresponding eigenfunctions blow up at $r=0$ with appropriate boundary conditions specified. For $d=2,3$, a family of self-adjoint operators exists; it is labeled by single parameter $\alpha$, related with the renormalized value of the coupling constant. This family describes the $s$-wave zero-range scattering, with the scattering length proportional to $-1/\alpha$. Unlike the one-dimensional case, where $\alpha$ represents the coupling constant of the $\delta$-interaction directly, the two and three-dimensional cases exhibit different behavior. Here, we focus on the case of a single point-like source at the origin and consider $d=3$. Our goal here is to express the vacuum polarization effects in terms of the scattering length.

\bigskip

The paper is structured as follows: The Introduction comes first, followed by a discussion on self-adjoint extensions for the Laplacian in three dimensions in Section~\ref{Self_adj}. Section \ref{Green} focuses on derivation of the renormalized Hadamard Green's function using a non-singular function with compact support, as proposed in \cite{Grats17, Grats20}. Direct mode summation (with the points-splitting procedure) is used to construct the Hadamard's function for the self-adjoint extension of the Hamiltonian. Section \ref{VEV} is devoted to computation of the renormalized vacuum averages $\langle \phi^{\kn
2}(x)\rangle_{\rm ren}$ and $\langle T_{\mu\nu}(x)\rangle_{\rm
ren}$ in the self-adjoint scheme. The last Section \ref{Concl} presents a comparison with results for massless field and offers a discussion and summary of the findings.
Two appendices describe transformations of  the main two-parameter integral and its asymptotic expansions with respect to both arguments and their product.

We use the units  $\hbar=c=1$; the metric signature is  $(+---)$. The notation $g_{\mu\nu}$ will denote the same Minkowski metric in any curvilinear coordinates.

\section{Self-adjoint extension of Laplacian}\label{Self_adj}

Written formally, the Klein\,--\,Gordon equation with  $\delta$-like potential reads
\begin{align}\label{Klein}
\Big(\kn\frac{\partial^2}{\partial t^2}-\Delta +m^2+\lambda\delta(\boldsymbol{x})\akn\Big)\kn \phi(t, \boldsymbol{x})=0  \,,
\end{align}
and the search of positive- and negative-frequency solutions,
\begin{align}
\phi_{\omega}^{(\pm)}(t, \boldsymbol{x})=\e^{\mp i\omega
t}\,u_{\omega}(\boldsymbol{x})\nn\,,
\end{align}
reduces to the possible  solutions to the equation
\begin{align}\label{Schro}
\big(-\Delta +\lambda\kn\delta(\boldsymbol{x})\big)\kn u_{\omega}(\boldsymbol{x})=(\omega^2-m^2) \kn u_{\omega}(\boldsymbol{x})\,.
\end{align}

Equation (\ref{Schro}) has a form of a stationary Schr\"{o}dinger equation for the quantum particle with mass $m=1/2$ in the field generated by zero-range potential.
However, this formally written equation is defined incompletely, since one should define the heuristically-written expression  $\lambda\kn\delta(\boldsymbol{x})$ for the potential.

It is believed that the correct description of zero-range (or $\delta^d$-like) interactions  for the spatial dimensionalities
 $d=1$, $d=
2$ and $d= 3$ is achieved by   construction of self-adjoint extension (SAE) of positively-defined operator $-\Delta$, acting on ${\mathbb{R}}^d\backslash\{\mathbf{0}\}$  \cite{Reed,Albeverio,Gitman}.
In the case
 \mbox{$d=1$} such a construction may be done by fixation of the value of a first-derivative jump for the wavefunction at a source's locus.
At that, the properly chosen parameter of extension coincides with the intensity $\lambda$ of one-dimensional
  $\delta$-interaction. In the case of two and three spatial dimensions the operator $-\Delta$ remains to be self-adjoint even if the eigenfunctions blow up at the potential locus.
  The proper condition consists in the requirement that eigenfunction has to be  locally integrable with square and to satisfy the boundary condition. For any of these two specified dimensionalities of a configurational space, one has a single-parameter family of self-adjoint operators. Notice, the SAE parameter already is not  connected directly  with the intensity   of
  $\delta^{d}$-interaction, but just determines the scattering length for the wave with vanishing angular momentum (so-called $s$-wave)
\cite{Reed,Albeverio}\,\,(see, also \cite{Jackiw}). For $d\geqslant 4$ the operator
$-\Delta$ is essentially self-adjoint itself, and the pointlike interactions can not be introduced in such a way.

 Therefore, in our problem we consider self-adjoint extension of the positively-defined operator  $H=-\Delta$,
acting on   ${\mathbb{R}}^3\backslash\{\mathbf{0}\}$. In spherical coordinates, the space $L^2(\mathbb{R}^3)$
can be decomposed into a direct sum
\begin{align}
L^2(\mathbb{R}^3)=\bigoplus\limits_{l=0}^{\infty}L^2\big((0, \infty);
r^2 dr\big)\bigotimes \big[Y_{l, -l}, \ldots, Y_{l, l}\big]\, ,
\end{align}
where  $L^2((0, \infty);
r^2 dr)$ denotes the space of functions locally integrable on  $\mathbb{R}_+$ with measure $r^2$, while $\{Y_{l, m}\mid l= 0, 1, 2,\, \pp\, ,\, m=0, \pm 1, \pm 2,
\ldots, \pm l\}$ ---   corresponding spherical harmonics, and $[\pp]$ means a linear span of corresponding vectors.

Respectively, the operator $H$  is decomposed into a sum
(see e.g. \cite{Reed,Albeverio,Gitman} and the literature therein)
\begin{align}
H=\bigoplus\limits_{l=0}^{\infty}H_l\bigotimes\mathbf{1}\, ,
\end{align}
where partial Hamiltonians are
\begin{align}
H_l=-\frac{d^{{\kern 0.5pt}2}}{dr^2}-\frac{2}{r}\frac{d}{dr}+\frac{l(l+1)}{r^2}\,
  ,\qquad l=0, 1, 2, \pp\, .
\end{align}

It is known that operators
 $H_l$ are self-adjoint itself for any $l\geqslant 1$, whereas the single-parameter family of self-adjoint extensions of $H_{0}$ is given by \cite{Albeverio,Gitman}
\begin{align}\label{hzero}
&H_{0, \alpha}=-\Delta_{0,
\alpha}=-\frac{d^{{\kern 0.5pt}2}}{dr^2}-\frac{2}{r}\frac{d}{dr}\, ,\\
&{\cal{D}}(H_{0, \alpha})=\Big\{u_{\alpha}\in L^2((0, \infty); r^2
dr)\, ;\,\, 4\pi\alpha\lim\limits_{r\to+ 0}r
u_{\alpha}(r)=\lim\limits_{r\to + 0}[u_{\alpha}+r
u'_{\alpha}]\Big\}\, ,\nn
\end{align}
where \mbox{$-\infty<\alpha\leqslant \infty$}. At that, the absence of interaction corresponds to the limiting case  \mbox{$\alpha =+\infty$}.

To conclude, the self-adjoint extension of the operator $H=-\Delta$
affects the $l=0$ state only and defines the zero-range interaction in  $s$-wave.

\section{Green's Function}\label{Green}

The eigenvalue/eigenfunction problem for the operator $H_{0,\alpha}$,
\begin{align}\label{Schro1}
H_{0,\alpha}\kn u_{p\alpha}(r)=p^2 u_{\omega\alpha}(r)\,,\qquad p^2=(\omega^2-m^2)\, ,
\end{align}
reduces to the study of equation
\begin{align}\label{Schro2}
  u''_{p\alpha}  +\frac{2}{r}\, u'_{p\alpha} +p^{\kn 2} u_{p\alpha} =0\, .
\end{align}
By the substitution
 $u_{p\alpha}=\chi_{p\alpha}(r)/r$
it becomes
$$ \chi_{p\alpha}''=-p^2 \chi_{p\alpha}\,,$$  which for real values of $p^2$ has the following solutions:
\begin{align}
 \chi_{p\alpha}=\left\{
                       \begin{array}{ll}
                          \mu_+\sin p r +\nu_+ \cos p r, & \hbox{$p^{\kn 2}>0$;} \\
                         \mu_- \e^{-|p|r} + \nu_- \e^{+|p|r} , & \hbox{$p^{\kn 2}<0$;} \\
                         \mu_0 r+\nu_0, & \hbox{$p=0$,}
                       \end{array}
                     \right.\nn
\end{align}
where coefficients are determined by the boundary conditions and by the requirement of local integrability with  square.
The latter fixes $ \nu_-= \mu_0=\nu_0=0$.

In result, for $p^2>0$  the $s$-wave solution differs from the regular one $\sim\sin p
r/r$, which corresponds to the empty Minkowski space with no interaction. It is given by
$$u_{p\alpha}(r) = C_{p\alpha}\Bigl(\frac{\sin p r}{r}+ \mathrm{tg}\kn\delta_{p\alpha}\,\frac{\cos p r}{r}\Bigr),$$
with some mixture angle $\delta_{p\alpha}$, which should be fixed from boundary conditions at $r=0$:
$$\mathrm{tg}\kn \delta_{p\alpha} = \frac{p}{4\pi\alpha}\,.$$
The coefficient $C_{p\alpha}$ is to be found from the normalization condition
$$i\int d  ^3{x}\,{\phi_{\,\omega}^{(\pm)^{\ast}}} \,\overleftrightarrow{{\partial}^{\vphantom{(\pm)}}_{\kn t}}\,
{\phi_{\,\omega^{\smash{\prime}}}^{(\pm)}}=\pm \delta(\omega-\omega')\,,$$
which is equivalent to the requirement
$$\int d^3{x}\,u^{\ast}_{p\alpha}\,
u^{\vphantom{\ast}}_{p'\alpha}=\frac{\delta(p-p')}{2p} \,.$$
In what follows $C_{p\alpha}= {\cos\delta_{p\alpha}}/({2\pi\sqrt{p}})\, .$

Therefore, the $s$-wave solution takes the form:
\begin{align}\label{s-wave}
u_{p\alpha}(r)=\frac{1}{2\pi r}\,\frac{\sin(p r +
\delta_{p \alpha})}{\sqrt{p}}\, .
\end{align}

Such a solution exists for any $p^2>0$ and for all $-\infty<\alpha\leqslant \infty$.

If $\alpha\geqslant 0$, then the self-adjoint operator
$H_{0, \alpha}$ has no negative eigenvalues, and the complete set of its eigenfunctions consists of
$\{u_{p l m}\}, l>0,$ and $u_{p\alpha}(r)$,
where $u_{p l m}$  is to be represented as solutions to
$$ u_{p l m}(r, \theta, \varphi) =\upsilon_{p l}(r ) Y_{lm}( \theta, \varphi)\,, \qquad  H_l \upsilon_{p l}=p^2\upsilon_{p l}\,.  $$

In the case
$-\infty<\alpha< 0$ the hamiltonian $H_{0, \alpha}$
has single discrete eigenvalue
$$p^2_{\alpha}=-(4\pi\alpha)^2\, ,\qquad \omega_{\alpha}^2=-(4\pi\alpha)^2+m^2\, ,$$
while the corresponding eigenfunction, normalized by $\langle u_{0, \alpha} \akn\mid \akn u_{0, \alpha}\rangle =1 $, equals
$$
u_{0, \alpha}(r)= \frac{\sqrt{-2\alpha}}{r} \,\e^{4\pi\alpha r}\,.
$$
In Quantum Mechanics, such a solution corresponds to the bound state of a particle in the background of a point-like {\it attractive} potential.
In our problem, for the case \mbox{$m^2<(4\pi\alpha)^2$}, the frequency square becomes negative, and the corresponding solution to the Klein\,--\,Gordon equation becomes unstable. In what follows, for the approach consistency, we should restrict our consideration  by the case $\alpha\geqslant 0$. In this setup, all the extensions of the $(-\Delta)$ operator are positively-defined self-adjoint operators, the spectrum of  $H_{0, \alpha}$ is continuous and covers the whole semi-axis
${\mathbb{R}}_{+}=[\,0, \infty)$, while instabilities are absent.

Let  $\{u_{p l  m}\}$ to be the complete set of eigenfunctions of the \textit{free} Laplacian.  Then the Hadamard Green's function may be represented as
\begin{align}
D_{\alpha}^{(1)}(x,
x')&=\frac{1}{2}\, \Big\langle\phi(x)\kn\phi(x')+\phi(x')\kn\phi(x)\Big\rangle_{\rm
vac}=\nn \\
&={\rm
Re}\int\limits_{0}^{\infty}d\omega\,e^{-i\omega(t-t')} \biggl[u^{\vphantom{\ast}}_{p
\alpha}(x)\,u^{\ast}_{p
\alpha}(x')+\sum\limits_{l=1}^{\infty}\sum\limits_{m=-l}^{l}  \! u^{\vphantom{\ast}}_{p
l m}(x)\,u^{\ast}_{p l m}(x')\biggr]\kn.
\end{align}
Notice for the formal identification $\alpha=\infty$ the latter coincides with the  Hadamard Green's function for the case of no interaction:
\begin{align}
D^{(1)}_{\infty}(x, x')={\rm
Re}\int\limits_{0}^{\infty}d\omega\,\e^{-i\omega(t-t')} \sum\limits_{l=0}^{\infty}\sum\limits_{m=-l}^{l}
u^{\vphantom{\ast}}_{\omega l m}(x)\,u^{\ast}_{\omega l m}(x')\,. \nn
\end{align}
This Green's function should determine the vacuum polarization effects in the Minkowski background: the vacuum expectation values for the field-square and for the Energy-Momentum tensor. In the no-interaction case, the latter ones are regarded to be vanishing.

In other words, $D^{(1)}_{\infty}(x, x')$ is to be regarded as ''zero-level'' Green's function, while the correction $D^{(1)}_{\rm ren}$, generated by the point interaction, is to be considered as renormalized
Hadamard Green's function:
$$D^{(1)}_{\rm ren}=D_{\alpha}^{(1)}-D^{(1)}_{\infty}\, . $$

In result, we find that, upon subtraction, all higher harmonics mutually cancel, and the renormalized Green's function is determined by the deformation of $s$-wave only:
\begin{align}
D^{(1)}_{\rm ren}(x, x')={\rm Re}\int\limits_{0}^{\infty}
d\omega\,\e^{-i\omega(t-t')}\,\Big[u^{\vphantom{\ast}}_{p\alpha}(x)\,u_{p
\alpha}^{\ast}(x')-u^{\vphantom{\ast}}_{p \infty}(x)\,u_{p
\infty}^{\ast}(x')\Big]\, . \nn
\end{align}
Here the  identification $u_{p 0 0} = u_{p\infty} $ was taken into account.

Substituting  $u_{p
\alpha}$ by eq.\,(\ref{s-wave}) and  carrying out the integration-variable change   $\omega \longrightarrow z=p/4\pi\alpha$, one finally obtains the renormalized
Hadamard Green's function in the form
\begin{align}\label{D1xx2}
D^{(1)}_{\rm ren}(x,
x')=\frac{1}{4\pi^2  rr'}\! \int\limits_{0}^{\infty} \!dz\,z\,\frac{\cos\akn \big[\sqrt{ (4\pi\alpha z)^2+m^2}(t-t')\akn \big]}{\sqrt{  z^2+(m/4\pi\alpha)^2}(1+z^2)}\,
\Bigl(\sin\akn \big[4\pi\alpha z (r+r')\big]\akn+ \akn z\,\cos\akn\big[4\pi\alpha z(r+r')\big]\akn\Bigr)\,.
\end{align}
In the massless limit, the latter coincides with the expression derived previously \cite{Grats2019,Graz21}.



\section{Vacuum Expectation Values}\label{VEV}

As it was mentioned, the basic quantities which indicate on the vacuum polarization, are the Field-Square $\langle \phi^{\kn 2}\rangle $ and the operator of  Energy-Momentum Tensor $\langle  T_{\mu\nu}\rangle $. We compute renormalized values of them
with help of the  renormalized
Hadamard Green's function $D^{(1)}_{\rm ren}(x,
x')$.

\subsection{Field Square}
In the coincident-point limit (CPL) we fix $t=t'$, $r'=r$, and the vacuum expectation value for the field square equals
\begin{align}
\langle\phi^2(x)\rangle^
{\vphantom{(\ast)}}_{\rm ren}=D^{(1)}_{\alpha}(x,
x)=\frac{1}{4\kn\pi^2 r^2}\,\mathcal{J}\Big(8\pi\alpha r,\frac{m}{4\pi\alpha}\Big) , \nn
\end{align}
where we introduce the following function of two arguments:
\begin{align}\label{girp4}
\mathcal{J}(\beta,a) \equiv\int\limits_{0}^{\infty} d z\,\frac{1}{1+z^2}\,\frac{z}{\sqrt{z^2+a^2}}
\Bigl(\sin\beta z+z\cos\beta z\Bigr)\, .
\end{align}
Transformations of this integral are considered in the Appendix~\ref{basint}.

 Introduce two quantities with dimension of length: the Compton length $l_c=m^{-1}$ and the scattering length $d_s=(4\pi \alpha)^{-1}$. In these notations,
 \begin{align}
\langle\phi^2(x)\rangle^
{\vphantom{(\ast)}}_{\rm ren}= \frac{1}{4\kn\pi^2 r^2}\,\mathcal{J}\Big(\frac{2r}{d_s},\frac{d_s}{l_c}\Big) . \nn
\end{align}

 We notice, that the product of two arguments  represents the main base of classification on limiting asymptotic regimes (Appendix \ref{asymint}). It equals $2r/l_c$, i.e. the  relative distance with respect to characteristic length scale of the field's mass.

The limit $d_s=0^+$ corresponds to the non-interacting hamiltonian. The limit $l_c\to \infty$ corresponds to the  massless field.
In this case all the results reproduce considered before \cite{Fermi1,Grats2019,Graz21}.

By physical reasons, we can readily investigate the directions of monotonic change of the renormalized $\langle \phi^2 \rangle$:
\begin{enumerate}
  \item for fixed $m$ and $d_s$, $\langle \phi^{\kn 2} \rangle_{\ren}$ monotonically decreases with growth of $r$, to  $\langle \phi^{\kn 2} (r\to \infty)\rangle_{\ren}=0$;
  \item  for fixed $r$ and $d_s$, $\langle \phi^{\kn 2}  \rangle_{\ren}$ monotonically decreases as $m$ grows, it follows immediately from generic form (\ref{girp4});
  \item   for fixed $r$ and $m$, $\langle \phi^{\kn 2}  \rangle_{\ren}$ monotonically decreases as $\alpha$  grows ($d_s$ falls), since from physical reasons the limit  $\alpha \to +\infty$ implies the absence of polarization effect.
\end{enumerate}

\medskip

\textbf{Asymptotic regimes.} The basic asymptotic regimes of $\mathcal{J}$ with abstract arguments $\beta$ and $a$ are considered in the Appendix \ref{asymint} in detail.

In the problem at hand we have two lengthy constants  $l_c=1/m$  and $d_s=1/4\pi \alpha$, and also the distance $r$ from the observation point to the  locus of potential.

For the massive field, $r\to \infty$ implies $r\gg \max\{l_c,d_s\}$. Here the first argument of $\mathcal{J}$ satisfies $\beta\gg 1$, while the second one is arbitrary. In this case we use asymptotic expansion
(\ref{macduck1}) of the Macdonald functions,
what yields the asymptotic
(\ref{blatt17}) for $\mathcal{J} $:
\begin{align}
 \mathcal{J}  \Big(\frac{2r}{d_s},\frac{d_s}{l_c}\Big) \simeq\frac{d_s}{d_s+l_c}\, K_0\Big(\frac{2r}{l_c}\Big) \kn.\nn
\end{align}

Hence
\begin{align}\label{phiw0}
\big \langle  \phi^{\kn 2}\big\rangle_{\rm
ren} \simeq \frac{ 1}{8 \pi^{3/2}  } \,  \frac{\sqrt{l_c} }{1+l_c/d_s } \,\frac{\e^{-2r/l_c}}{r^{5/2}}\,.
\end{align}

Then for $l_c\ll d_s \ll r$ (ultra-massive field) one obtains
\begin{align}\label{phiw1}
\big \langle  \phi^{\kn 2}\big\rangle_{\rm
ren} \simeq \frac{ 1}{8 \pi^{3/2}  } \,  \frac{\sqrt{l_c} }{r^{5/2}} \,\e^{-2r/l_c}\,.
\end{align}

For $d_s\ll l_c \ll r$ (low values of $d_s$, where the interaction is small), $\phisq$ is given by
\begin{align}\label{phiw2}
\big \langle  \phi^{\kn 2}\big\rangle_{\rm
ren} \simeq \frac{ 1}{8 \pi^{3/2}  } \,  \frac{d_s}{\sqrt{l_c} } \,\frac{\e^{-2r/l_c}}{r^{5/2}}\,.
\end{align}
In all the regimes above, the effect decreases exponentially (as $r$ grows) like $\exp(-2r/l_c)$.

Also, there is a special regime $ d_s \ll r \leqslant l_c$ (almost massless case). Making use of (\ref{blatt2kk}), one obtains
\begin{align}\label{phiw4}
\big\langle \phi^{\kn 2}\big\rangle _{\rm ren}=
\frac{ 1}{4\pi^2 r^2}  \frac{d_s}{l_c} \, K_1 \Big( \frac{2r}{l_c}\Big)
\Big[
1+ \mathcal{O}\Big( \frac{d_s}{ r }\Big)\Big]\,.
\end{align}

Finally, for the massless field,
large distances imply $r\gg d_s$ and thus $\beta\gg 1$.  When $l_c\to +\infty$, we can directly apply the limit $\hat{K}_1(0):=[x\kn {K}_1(x)]_{x\to 0^+}=1$ to (\ref{phiw4}), to get
\begin{align}\label{phiw3}
\big\langle \phi^{\kn 2}\big\rangle _{\rm ren}=
\frac{ d_s}{8\pi^2 r^3}\kn \,
\Big[
1+ \mathcal{O}\Big( \frac{d_s}{ r }\Big)\Big]\,.
\end{align}
The same concerns the transition case $d_s\leqslant r \ll l_c$.

\medskip

Now consider small distances.
 Here we use small-argument expansion (\ref{macduck3}) of the Macdonald functions. For the case  $\{r\ll l_c \leqslant d_s\}$ we make use of
(\ref{blatt4cf123})
and thus for $\langle \phi^{\kn 2}  \rangle_{\ren}$ the asymptotic
\begin{align}\label{phiw7}
\big\langle  \phi^{\kn 2}\big\rangle _{\rm ren}=
\frac{ 1}{4\pi^2 r^2}\,\Big[\ln \frac{l_c}{r}+ \mathcal{O} (1 )\Big]
\end{align}
holds.

For   ultra-large or infinite Compton length  $r\ll d_s \leqslant l_c$, then one uses (\ref{blatt4uu}),
\begin{align}\label{phiw8}
\big\langle  \phi^{\kn 2} \big\rangle_{\rm ren}=
 \frac{1}{4\pi^2r^2}  \Big[\ln \frac{d_s}{r}+\mathcal{O}(1)\Big]\kn.
\end{align}

For the transition case $l_c \leqslant r\ll d_s $ we use the asymptotic (\ref{blatt2yyy}), that yields
\begin{align}\label{phiw9}
\big\langle \phi^{\kn 2}\big\rangle _{\rm ren}=
\frac{ 1}{4\pi^2 r^2}\kn \,K_0\Big( \frac{2r}{l_c}\Big)
\Big[
1+ \mathcal{O}\Big( \frac{r}{d_s}\Big)\Big]\,.
\end{align}
For the limit $l_c\ll r$ the latter simplifies to
\begin{align}\label{phiw10}
\big \langle  \phi^{\kn 2}\big\rangle_{\rm
ren} \simeq \frac{ 1}{8 \pi^{3/2}  } \,  \frac{\sqrt{l_c} }{r^{5/2}} \,\e^{-2r/l_c}\,;
\end{align}
the same as in (\ref{phiw1}). Therefore, (i) if $d_s\gg l_c$ then the exponential decay with the main asymptotic (\ref{phiw1}) takes place independently of the mutual relation of $r$ versus $d_s$;
(ii) the exponential decay $\e^{-2r/l_c}$ happens at $r>l_c$ independently of the value $d_s$. However, such an exponential fall was predicted for the massive scalar field.

\begin{figure}
 \begin{center}
\includegraphics[width=10cm]{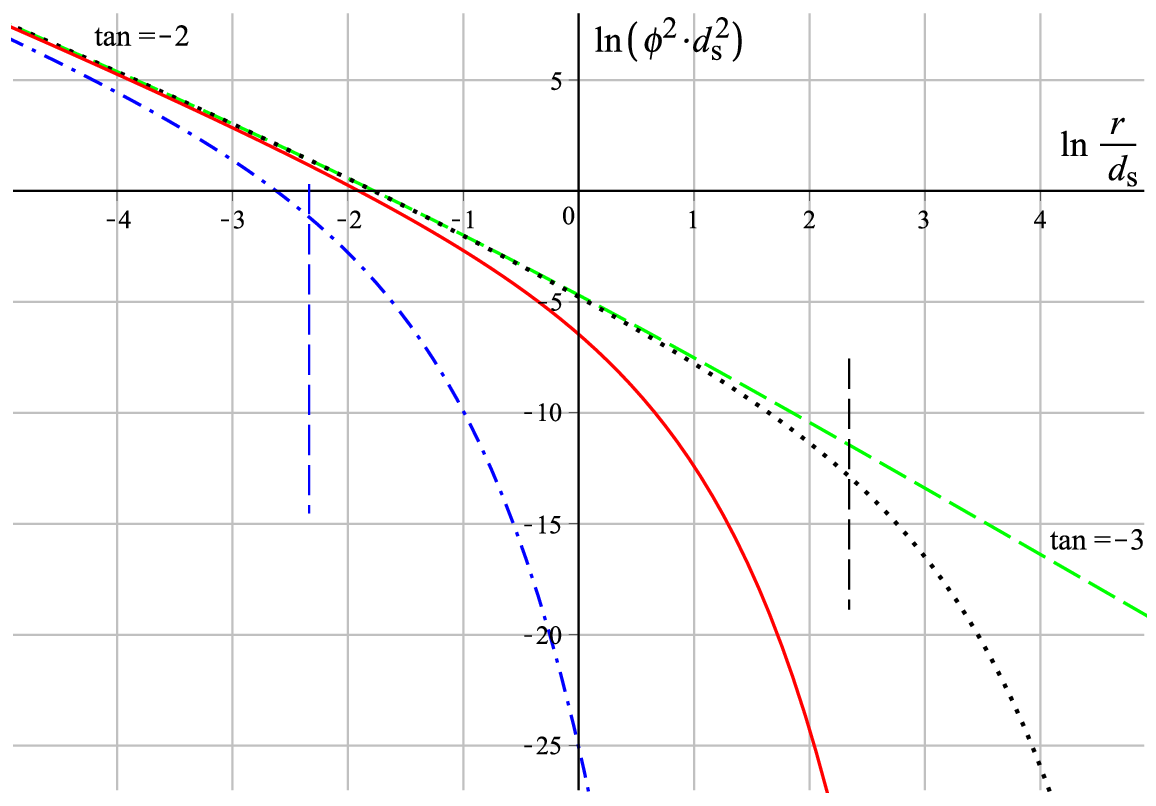}
 \caption{Renormalized field-square  in doubly logarithmic scale: for massless field (green dashed), for $l_c/d_s=10$ (black dotted), $l_c/d_s=1$ (red solid) and $l_c/d_s=0.1$ (blue dashdotted). The value $r=l_c$ is marked by dash of corresponding color. The value $r=d_s$ corresponds to the ordinate-axis for each curve.}
 \label{Philog}
\end{center}
\end{figure}

\medskip

\textbf{Qualitative conclusion.} Combining the information on the behavior of  $\langle \phi^{\kn 2}  \rangle_{\ren}$ in asymptotic regimes, we can conclude that qualitatively the behavior of field-square in the entire domain of $r$ drastically depends upon the relation between two lengthy constants:
\begin{enumerate}
  \item  \mbox{$\boldsymbol{l}_{\boldsymbol{c}}\boldsymbol{=}\boldsymbol{\infty}:$} here at $r<d_s$ the behavior is $\langle \phi^{\kn 2}  \rangle_{\ren}\sim  r^{-2}\ln (d_s/r)$; in a some transition domain $T_s$  (where $r \sim d_s$)   $\langle \phi^{\kn 2}  \rangle_{\ren}$ goes like $ r^{-(2+\epsilon)}$ (where $0<\epsilon<1$ increases), and  at $r\gg d_s$ we have $\langle \phi^{\kn 2}  \rangle_{\ren}\sim r^{-3}$;
  \item  \mbox{$\boldsymbol{d}_{\boldsymbol{s}}\boldsymbol{\ll} \boldsymbol{l}_{\boldsymbol{c}}:$} here at \mbox{$r<d_s$} one has  $\langle \phi^{\kn 2}  \rangle_{\ren}\sim  r^{-2}\ln (d_s/r)$; in a wide middle domain \mbox{$r\in  [\sim d_s, \sim l_c]$}    $\langle \phi^{\kn 2}  \rangle_{\ren}$ goes like $ r^{-3}$, and  at \mbox{$r>l_c$} --- exponential decay \mbox{$\langle \phi^{\kn 2}  \rangle_{\ren}\sim  \exp (-2r/l_c)$}. Also, there are two transition domains: $T_s$ described above, and $T_c$ , where \mbox{$r \sim l_c$} and the function rapidly changes the power-like decay to the exponential one;
  \item   \mbox{$\boldsymbol{l}_{\boldsymbol{c}}\boldsymbol{\lesssim} \boldsymbol{d}_{\boldsymbol{s}}:$} here at \mbox{$r<l_c$} the field square \mbox{$\langle \phi^{\kn 2}  \rangle_{\ren}\sim  r^{-2}\ln (l_c/r)$}; in   transition domain  $T_c$   it changes behavior, while in  \mbox{$r>l_c$} domain \mbox{$\langle \phi^{\kn 2}  \rangle_{\ren} $} exponentially decays \mbox{$\sim  \exp (-2r/l_c)$}.
\end{enumerate}
To illustrate it, the plot of $\langle \phi^{\kn 2}  \rangle_{\ren} $ versus $r/d_s$ is shown on the Fig.\,\ref{Philog} in doubly logarithmic mode.

Finally, we analyze  dependence of  $\langle  \phi^{\kn 2}\rangle_{\rm ren}$ with respect to the SAE parameter $\alpha$ (with fixed $r$ and $m$).
In contrast with two other lengthy quantities, the scattering length $d_s$ appears in both arguments of $\mathcal{J}$, so that from their product $d_s$ disappears.
In order to reveal the asymptotic behavior (at large and small $d_s$), with no loss of generality, we can use asymptotic Case\,III described in the Appendix \ref{asymint}, for any mutual relation between $r$ and $l_c$.

As $d_s\to 0^+$, the result is given by eq.\,(\ref{phiw4}); in the opposite case ($d_s\to \infty$) the asymptotic is given by the
expansion (\ref{blatt2yyy}). For the latter, the renormalized $\langle  \phi^{\kn 2}\rangle$ has a maximal value (horizontal asymptote) $\langle  \phi^{\kn 2}\rangle_{\rm max}$ given by  eq.\,(\ref{phiw9}).

In what follows, introducing (for confidence level $p\lesssim 1$) the lengthy parameter $ d_s^{\kn(p)}$
 as  $$ \mathcal{J}\bigg(\frac{2r}{ d_s^{\kn(p)}},\frac{ d_s^{\kn(p)}}{l_c} \bigg) = p\kn K_0\Big( \frac{2r}{l_c}\Big)\kn,  $$
it takes the form
 \begin{align}\label{phiw11}
 d_s^{\kn(p)}\sim \frac{1}{1-p} \,f_p\Big(\frac{r}{l_c}\Big)\kn l_c\,,
\end{align}
 where $f_p(\cdot)$ is an increasing slowly altering function of order of unity.

\begin{figure}
 \begin{center}
\includegraphics[width=10cm]{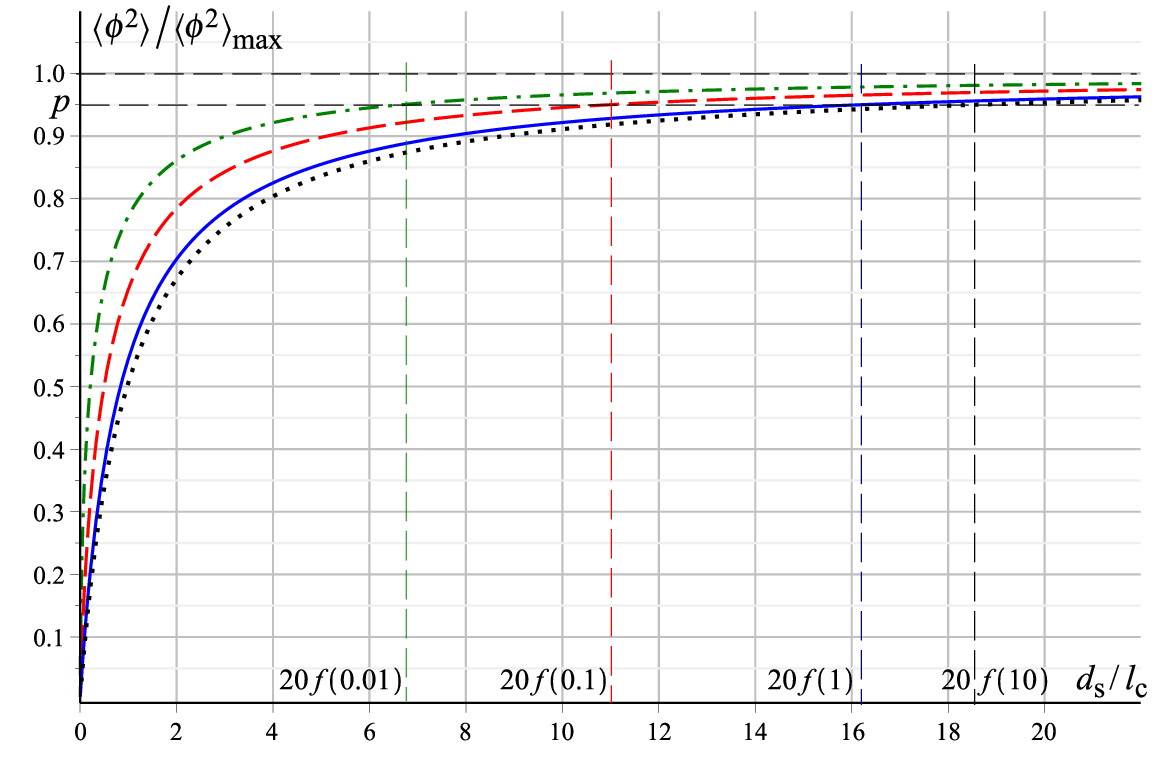}
 \caption{Vacuum mean of field-square, normalized by $\langle  \phi^{\kn 2}\rangle_{\rm max}$ as a function of $d_s$ (in units $l_c=1$): for $r/l_c=0.01$ (green dashdotted), for $r/l_c=0.1$ (red dashed), $r/l=1$ (blue solid) and $r/l_c=10$ (black dotted). The values $ d_s^{\kn(p=0.95)}$ are marked by vertical dash lines of corresponding color. The corresponding values of $f_{(p=0.95)}$ are: $f_{0.95}(0.01)=0.333$,  $f_{0.95}(0.1)=0.550$, $f_{0.95}(1)=0.810$, $f_{0.95}(10)=0.928$, $f_{0.95}(100)=0.948$.}
 \label{phi_ds}
\end{center}
\end{figure}

 The corresponding plots of  $\langle  \phi^{\kn 2}\rangle_{\rm ren}$ (normalized by  $\langle  \phi^{\kn 2}\rangle_{\rm max}$) versus $d_s$ are presented on the Fig.\,\ref{phi_ds}.

 Thereby, for large values of the scattering length, $d_s>d_s^{\kn(p)}$, the vacuum polarization becomes insensitive to it and goes like a constant.

\subsection{Renormalized Energy-Momentum Tensor}

The Energy-Momentum Tensor (EMT) of the real massive scalar field in a flat four-dimensional space-time is given by
\begin{align}\label{TEI}
 T_{\mu}^{\nu} = (1-2\xi)\,\partial^{\nu}\phi\,\partial_{\mu}\phi+\Big(\frac{1}{2}-2\xi\Big)
\delta_{\mu}^{\nu}\big(m^2\phi^{\kn 2}-\partial^{\lambda}\phi\,\partial_{\lambda}\phi\big)
-2\xi\phi \nabla^{\nu}\partial_{\mu}\phi +\frac{1}{2}\,\xi\kn\delta_{\mu}^{\nu} \big( \phi \,\square \phi  +m^2 \phi^{\kn 2}\big)\,,
\end{align}
where $\square:=g^{\kn \mu\nu}\nabla_{\akn\mu}\nabla_{\akn\nu}$.

Since $\phi\kn$ satisfies the homogeneous Klein\,--\,Gordon equation, the last term may be omitted. Then the renormalized EMT operator is determined by the appropriate differentiations of the renormalized Hadamard Green's function $  D^{(1)}_{\alpha}(x,x')  $:
\begin{align}\label{Tmunu3}
 \big\langle T_{\mu}^{\nu}\big\rangle_{\rm ren}=\lim\limits_{x'\rightarrow
x}\left[(1-2\xi)\,\partial^{\nu}\partial'_{\mu}+\Big(\frac{1}{2}-2\xi\Big)
\delta_{\mu}^{\nu}\left(m^2-\partial^{\lambda}\partial'_{\lambda}\right)
-\xi\left(\nabla^{\nu}\partial_{\mu}+{\nabla'}^{\nu}\partial'_{\mu}\right)\right]D^{(1)}_{\alpha}(x,
x')\,.
\end{align}
Notice, the covariant derivatives are essential in spherical coordinates, where the following non-vanishing Christoffel symbols are useful: $\Gamma_{\theta\theta}^r=-r$ and $\Gamma_{ \varphi\varphi}^r=-r\sin^2 \theta$.

Upon differentiation (with respect to
 $t(t')$ and $r(r')$), the integrand in (\ref{D1xx2}) accumulates the integration variable   $z$ in   numerator, and thus at second differentiation the integral becomes UV-divergent. At the other hand, our basic integral $\mathcal{J}(\beta,a )$ is defined already at the coincident-point limit.
  In other words, we need the correspondence rules
for the derivatives of $D^{(1)}_{\alpha}(x,
x')$ (with respect to $t$, $t'$, $r$ and $r'$) and with subsequent CPL, and derivatives of $\mathcal{J}(\beta,a )$ with respect to $\beta$.

Consider first the derivatives over $r$: denote the integral in eq.\,(\ref{D1xx2}) as
\begin{align}
 & D^{(1)}_{\alpha}(x,
x')=\frac{I(t-t',r+r')}{4\pi^2 \,rr'}  \nn \\
&I := \int\limits_{0}^{\infty} dz\,\frac{\cos\akn \big[\sqrt{ (4\pi\alpha z)^2+m^2}(t-t')\akn \big]}{1+z^2}\,
\Bigl(\sin \akn\big[\kn 4\pi\alpha z (r+r')\big]+ z\, \cos\akn\big[\kn 4\pi\alpha z(r+r')\big]\Bigr)\,. \nn
\end{align}
Then in the CPL one obtains
\begin{align}\label{IlimJ}
I(0,2r)= \mathcal{J}\Big(8\pi \alpha r, \frac{m}{4\pi \alpha} \Big)\,.
\end{align}

Since both radial variables appear in the integrand of  $I$ as a sum, then
$$  \frac{\partial I(t-t',r+r')}{\partial r}= \frac{\partial I(t-t',r+r')}{\partial r'} = \frac{\partial I(t-t',r+r')}{\partial (r+r')}\,. \nn $$
Therefore, if to introduce new variable
$R\equiv r+r'$, then one notices that the limit  $r=r'$ does not change the total $R$, while the limit  $t=t'$ does not deal with the differentiation variable $R$ and just transforms $I$ into $\mathcal{J}$. In other words, two operations --- differentiation with respect to $R$ and taking the CPL in $I$ --- commute. Hence one can immediately differentiate   (\ref{IlimJ}) in the form
$$I(t-t,R)= \mathcal{J}\Big(8\pi \alpha r, \frac{m}{4\pi \alpha} \Big)\nn $$
with respect to  $R$  and thus, to get the correspondence rules for derivatives over radial coordinates:
$$ \frac{\partial I(t-t',r+r')}{\partial r}\Big|_{x'=x}= \frac{1}{2} \, \frac{\partial \mathcal{J}(8\pi \alpha r,a)}{\partial r}\,, \qquad\qquad  \frac{\partial^{{\kern 0.5pt} 2} I(t-t',r+r')}{\partial r^2}\Big|_{x'=x}= \frac{1}{4} \, \frac{\partial^{{\kern 0.5pt} 2} \mathcal{J}(8\pi \alpha r,a)}{\partial r^2}\,.\nn$$

The first temporal derivatives are proportional to $\sin (t-t')$ and thus vanish in the CPL:
$$ \frac{\partial I(t-t',r+r')}{\partial t}\Big|_{t'=t}=-\frac{\partial I(t-t',r+r')}{\partial t'}\Big|_{t'=t}=0\,.\nn $$

For the computation of the second temporal derivatives, one can use the Klein\,--\,Gordon equation. At the other hand, one can notice that constituents
of the integrand in $I $ satisfy the equation
$$ \left(c^2\frac{\partial^{{\kern 0.5pt} 2}}{\partial t^2}-\tilde{c}^2\frac{\partial^{{\kern 0.5pt} 2}}{\partial r^2}  \right) \! \!\bigg\{ \!
\begin{array}{c}
  \cos\left[\kn \tilde{c} \kn(t-t') \right]\sin \left[\kn c \kn(r+r')\right]  \\[3pt]
  \cos\left[\kn \tilde{c}\kn (t-t')\right]\cos \left[\kn c  \kn(r+r')\right]
\end{array}\!
  \bigg\} =0\nn
 $$
(where $c= 4\pi\alpha z$, $\tilde{c}= \sqrt{(4\pi\alpha z)^2+m^2}$) separately, as well as total integral $I $. The same concerns
operator $\big( c^2 {\partial }/\partial t\cdot  {\partial }/\partial t'+ \tilde{c}^2\partial/\partial r\cdot  \partial/\partial r'\big)$.

The derivative operators with respect to  $t$ and $r$ can be pushed outside the  integral. Therefore,
$$  \frac{\partial^{{\kern 0.5pt} 2}}{\partial t^2}=\frac{\tilde{c}^2}{c^2}   \frac{\partial^{{\kern 0.5pt} 2}}{\partial r^2}= \Big(1+\frac{m^2}{(4\pi \alpha z )^2}\Big)   \frac{\partial^{{\kern 0.5pt} 2}}{\partial r^2}\,,\nn$$
and thus the double $t$-differentiation is equivalent to the placement of the latter operator into the integrand. The first term of it yields the direct correspondence rule for differentiation of  $I$ and $\mathcal{J}$. In the second term we directly differentiate and take CPL, to obtain finally
$$  \frac{\partial^{{\kern 0.5pt} 2} I(t-t',r+r')}{\partial t^2}\Big|_{x'=x}=
\bigg(\frac{1}{4}\frac{\partial^{{\kern 0.5pt} 2}}{\partial r^2}-m^2\bigg)\mathcal{J}\,.\nn$$

Thereby, all the necessary derivatives of $I$ in the CPL may be transformed into the corresponding derivatives of $\mathcal{J}$ with respect to $r$.
Since $r$ appears in the first argument of $\mathcal{J}$ only, introduce the following combinations:
\begin{align}
\mathcal{J}_1(\beta,a)  := \beta \frac{\partial\mathcal{J}(\beta,a)}{\partial\beta}\,,\qquad\qquad
\mathcal{J}_2(\beta,a) := \beta^2 \frac{\partial^{{\kern 0.5pt} 2}\akn \mathcal{J}(\beta,a)}{\partial\beta^2} \,.\nn
\end{align}

In terms of introduced  $\mathcal{J}_k$'s, we compute all the components of renormalized EMT: among them, the non-vanishing ones are:
\begin{align} \label{Trr33}
\big\langle T_t^t \big\rangle_{\rm
ren}&  =\frac{1}{ 4\kn\pi^2 r^4}\left[\Big(\frac{1}{2}-2\xi+ m^2r^2  \Big)\mathcal{J}+\Big(2\xi- \frac{1}{2}  \Big)\mathcal{J}_1 - \xi  \mathcal{J}_2  \right]  \nn\\
 \big\langle T_r^r\big\rangle _{\rm
ren}&  =\frac{1}{ 4\kn\pi^2 r^4}\left[\Big(4\xi- \frac{1}{2}  \Big)\mathcal{J}+\Big(\frac{1}{2}  -2\xi \Big)\mathcal{J}_1   \right] \nn \\
 \big\langle T_\theta^\theta\big\rangle_{\rm ren} & =\frac{1}{ 4\kn\pi^2 r^4}\left[\Big(\frac{1}{2}-4\xi \Big)\mathcal{J}+\Big(3\xi- \frac{1}{2}  \Big)\mathcal{J}_1 +\Big(\frac{1}{4}- \xi \Big)\mathcal{J}_2  \right]  \nn \\
& = \big\langle T_\varphi^\varphi\big\rangle_{\rm
ren}\,,
\end{align}
where all $\mathcal{J}_k$ are to be evaluated at arguments
\begin{align}
\mathcal{J}_k=\mathcal{J}_k\Big(8\pi \alpha r,\frac{m}{4\pi \alpha}\Big) \,.\nn
\end{align}

The trace of EMT equals
\begin{align}\label{traccia}
\mathrm{Sp}\big \langle\kn  T \big\rangle_{\rm
ren}=\frac{1}{8 \pi^2 r^4} \Big[  (1-6 \xi)\Big(2\mathcal{J} -2\mathcal{J}_1 +\mathcal{J}_2 \Big)+2m^2r^2\mathcal{J} \Big]\, .
\end{align}
In the massless case and for the conformal field   ($\xi=1/6$) the trace vanishes, in accordance with theory \cite{Birrell, christ76}. Therefore, the conformal anomaly in the problem-at-hand is absent.

In the Appendix \ref{basint} it is shown that  $\mathcal{J}_{1,2}$ can be reduced into the superposition of  $\mathcal{J}$ and Macdonald functions $\hat{K}_{0,1}$ (the modified Bessel functions of 3rd kind). Hence the non-vanishing (diagonal) components of the renormalized EMT take the form
\begin{align}\label{pelad0}
\big \langle T_\nu^\nu\big\rangle_{\rm
ren}=\frac{1}{4 \pi^2 r^4} \left[A_{\nu,-1} \mathcal{J}\Big(\frac{2r}{d_s},\frac{d_s}{l_c}\Big)  +A_{\nu,0} {K}_0\Big(\frac{2r}{l_c}\Big)+ A_{\nu,1} \hat{K}_1\Big(\frac{2r}{l_c}\Big) \right]
\end{align}
(no summation over $\nu$). The values of coefficients $A_{\nu,\sigma}$ are presented in the Table\,\ref{tabcoef}.

\begin{table}[h]
\begin{center}\small \erh=7pt \tcs=0mm
\begin{tabular}{c p{1cm} c>{\hspace*{7.5mm}}c>{\hspace*{6mm}}c}
\toprule
  \multirow{2}*{
  \begin{tabular}{l}
 \footnotesize Index $ \nu$\\[-12pt]
   \footnotesize of the diagonal \\[-12pt]
  \footnotesize  component
  \end{tabular}
 } & & \multicolumn{3}{c}{\footnotesize Index $\sigma$} \\\cline{3-5}
  & &\footnotesize $ -1$ & \footnotesize $0$ &\footnotesize $ 1$
  \\  \midrule
\addlinespace[1.1pt]
  $t$ & & $\ds \Big(2\xi -\frac{1}{2}\Big) \Big(\frac{2r}{d_s}-1\Big)+\frac{r^2}{\smash{l_c^2}}-4\xi\frac{r^2}{\smash{d_s^2}} $ & $\ds-4\xi\frac{r^2}{\smash{l_c^2}}$& $\ds \frac{1}{2}+\xi \Big( \frac{2r}{d_s}-3\Big) $  \\[7pt]
\rowcolor{superlight-gray}
$r$ & & $\ds 4\xi -\frac{1}{2}+ (1-4\xi)\frac{r}{d_s}  $ & $\ds 0$ & $ \ds 2\xi -\frac{1}{2}$  \\[7pt]
$\theta,\lefteqn{ \varphi}$ & & $\ds\frac{1}{2}-4\xi + (6\xi-1)\frac{r}{d_s}+ (1-4\xi)\frac{r^2}{\smash{d_s^2}} $ & $ \ds (1-4\xi)\frac{r^2}{\smash{l_c^2}} $ & $\ds  \frac{3}{4}-4\xi+ \Big(2\xi-\frac{1}{2}\Big)\frac{r}{d_s}$   \\[6pt]\hline
\rowcolor{superlight-gray}
$\,\mathrm{Sp}$ & & $\ds (1-6\xi)  \Big(1-\frac{2r}{d_s\vphantom{q_{q_q}}}+\frac{2r^2}{\smash{d_s^2}}\Big)+\frac{r^2}{\smash{l_c^2}}$ & $ \ds 2 (1-6\xi) \frac{r^2}{l_c^2}  $ & $\ds \ds(1-6\xi)  \Big(\frac{3}{2}-\frac{r}{d_s}\Big)  $   \\
\addlinespace[-1.6pt]
  \bottomrule
\end{tabular}\caption{Coefficients $A_{\nu,\sigma}$ for the EMT in eq.\,(\ref{pelad0}), and for its trace.}\label{tabcoef}
\end{center}
\end{table}

\medskip

\textbf{Conservation.} Check, that the renormalized EMT, which was derived with help of the renormalized  Hadamard Green's function  $D^{(1)}_{\alpha}(x,
x')$, conserves:
\begin{align}\label{EMTcons}
\nabla_{\nu} \big\langle T_{\mu}^{\nu}\big\rangle_{\rm ren}=0\, .
\end{align}
It is known that the covariant divergence of symmetric two-rank tensor with mixture-typed indices reduces to
$$\nabla_{\nu} \big\langle T_{\mu}^{\nu}\big\rangle_{\rm ren} = |g|^{-1/2}\partial_{\nu} \Big(|g|^{ 1/2} \big\langle T_{\mu}^{\nu}\big\rangle_{\rm ren}\Big)-\frac{1}{2}\, \big\langle T^{\nu}_{\rho}\big\rangle_{\rm ren}g^{\lambda \rho} \partial_{\mu}g_{\nu\lambda}\,,\nn  $$
hence for free indices $\mu=t, \theta, \varphi$ it holds trivially, since EMT depends upon the radial coordinate only.
For the component $\mu=r$ the conservation is equivalent to the validity of identity
\begin{align}\label{conservr}
\beta \frac{\partial R(\beta,a)}{\partial\beta}  -2\Big[R(\beta,a)+\Phi(\beta,a)\Big]=0\,,
\end{align}
where it was denoted $R :=    r^4 T_r^r$ and  $ \Phi :=   r^4 T_{\varphi}^{\varphi}$. It is useful to take the components of EMT in the form   (\ref{Trr33}); thus by direct computation one convinces oneself that
 (\ref{conservr}) is valid in terms of functions $\mathcal{J}_k$.

\medskip

\textbf{Asymptotic regimes.} Now we can rearrange the asymptotic regimes in accord with the information, got from the computation of $\langle  \phi^{\kn 2}\rangle_{\rm ren}$.

\medskip

\il For $d_s \ll l_c\ll 1$ we have to take into account corrections, since we do not know whether $a\beta=2r/l_c\gg1$ larger than $1/a=l_c/d_s\gg1$ or not.
From (\ref{blatt17nn}) we have
\begin{align}
 \mathcal{J}  =  \sqrt{\frac{\pi l_c}{4 r }}\,\e^{-2r/l_c}\bigg[ \frac{d_s}{l_c}+  \Big(\frac{3 }{16}\frac{d_s}{r}  -  \frac{d_s^2}{l_c^2}\Big)+
  \Big( \frac{d_s^3}{l_c^3}-\frac{7 }{16}\frac{d_s^2}{r l_c}  - \frac{15 }{512}\frac{d_s l_c}{r^2}\Big)+
 \pp   \bigg]\kn.\nn
\end{align}
After routine but straightforward computation
 we get
\begin{align}\label{Ttt0}
\big\langle T_{\mu}^{\nu}\big\rangle _{\rm ren}=
\frac{1-4\xi}{8\pi^{3/2}}\frac{d_s}{  l_c^{5/2} r^{5/2}}\,\e^{-2r/l_c}\mathrm{diag}\bigg(1+\frac{1+4\xi}{1-4\xi}\frac{l_c^2}{r d_s},0,1,1\bigg)\kn.
\end{align}
 Hereafter in the list of diagonal components we assume the variable sequence $t \prec r \prec \theta \prec \varphi$.

\medskip

\il For the case $r \gg l_c$, $d_s \gg l_c $, which covers cases \mbox{$l_c \ll d_s \ll r$} and \mbox{$l_c \ll r \ll d_s$} (compare eq.\,(\ref{phiw1}) with eq.\,(\ref{phiw10})),
we have $$\mathcal{J}  \Big(\frac{2r}{d_s},\frac{d_s}{l_c}\Big) \simeq  K_0\Big(\frac{2r}{l_c}\Big) \kn.$$
Therefore, for the basic asymptotic we have
\begin{align}\label{Ttt1}
\big\langle T_{\mu}^{\nu}\big\rangle _{\rm ren}=
\frac{1-4\xi}{8\pi^{3/2}}\frac{\e^{-2r/l_c}}{ l_c^{3/2} r^{5/2}}\,\mathrm{diag}\big(1,0 ,1,1 \big)\kn,
\end{align}
where we keep the same-order quantities for different components. In what follows, $ \big|\big\langle T_{r}^{r}\big\rangle\big| $ does not vanish but it is subleading with respect to $\big|\big\langle T_{t}^{t}\big\rangle\big|$.

\medskip

\il For the transition case $d_s \ll r \leqslant l_c$
we use expansion (\ref{blatt2kk}). Since any integer-index Macdonald function can be expressed via superposition of $K_0$ and $K_1$, it is more useful to do it, in order to combine them with those ones of
(\ref{pelad0}):
\begin{align}\label{blatt2kk2}
\mathcal{J} (\beta,a)=  a \kn K_1 (a\beta)- a^2   \bigg(K_0 (a\beta)+\frac{K_1(a\beta)}{a\beta}\bigg)+a^3\bigg( K_1 (a\beta)+\frac{K_0(a\beta)}{\alpha\beta}+\frac{2K_1(a\beta)}{\alpha^2 \beta^2}\bigg) + \mathcal{O}(a^4) \,,
\end{align}
to obtain for diagonal components:
\begin{align}\label{Ttt2}
\big\langle T_{ \nu}^{\nu}\big\rangle _{\rm ren}=
\frac{1}{4\pi^2 r^4}\, \frac{d_s}{l_c} \bigg[B_{\nu1}  K_1 \Big( \frac{2r}{l_c}\Big) + B_{\nu2} \frac{r }{l_c} \, K_0 \Big( \frac{2r}{l_c}\Big) + B_{\nu3}\kn \frac{r^2 }{l_c^2} \kn K_1 \Big( \frac{2r}{l_c}\Big)\bigg]\,,
\end{align}
where for simplicity we list $B'$s just in the text: $B_{t}=(1-6\xi,1-6\xi,1-4\xi)$, $B_{r}=(6\xi-1,4\xi-1,0)$, $B_{\theta}=B_{\varphi}=(3/2-9\xi,3/2-8\xi,1-4\xi)$.

In the limiting case $d_s \ll r \ll l_c$,  the dominance comes from the terms with minimal power of the Compton length in denominators:
\begin{align}\label{Ttt4}
\big\langle T_{ \mu}^{\nu}\big\rangle _{\rm ren}=
\frac{ 1-6 \xi}{4\pi^2 r^4}\,  \frac{d_s}{l_c} \, K_1 \Big( \frac{2r}{l_c}\Big)
\,\mathrm{diag}\Big(1,-1 ,\frac{3}{2},\frac{3}{2} \Big)\,.
\end{align}

Tending $l_c\to+\infty$ we get large-distance ($r\gg d_s$) asymptotic for the massless field:
\begin{align}\label{Ttt5}
\big\langle T_{ \mu}^{\nu}\big\rangle _{\rm ren}=
\frac{ 1-6 \xi}{8\pi^2 r^5}\,  d_s
\,\mathrm{diag}\Big(1,-1 ,\frac{3}{2},\frac{3}{2} \Big)\,, \qquad\qquad \xi \neq  \frac{1}{6}\,.
\end{align}

The two latter formulae represent the basic term for non-conformal coupling. For the conformal one we need in the next terms of expansion (\ref{blatt2kk2}).
At the other hand, the conformal coupling is actual for the massless field only. In this case we can start with the initially-massless field and make use
of $$ \mathcal{J} (\beta,0)= E_1(\beta)\,\exp(\beta)\,.$$
With help of asymptotic expansion (\ref{Ei_asy}), one deduces:
\begin{align}\label{Ttt65}
\big\langle T_{ \mu}^{\nu}\big\rangle _{\rm ren}=
\frac{ d_s^{\kn 2}}{32\pi^2 r^6}
\,\mathrm{diag}\Big(1,\frac{1}{3} ,-\frac{2}{3},-\frac{2}{3} \Big)\,, \qquad\qquad \xi = \frac{1}{6}\,.
\end{align}
The trace vanishes, indeed. Therefore, the conformal field decays faster.

\medskip

\il The case of small distances implies $r \ll \min\{ l_c, d_s \}$.

Making the comparison of (\ref{phiw7}) and (\ref{phiw8}), we see that qualitatively, the dependence is expected to be logarithmic, with the appropriate  log-scale factor $L$ to be $L=\min\{ l_c, d_s \}$. Let us verify it.

From eq.\,(\ref{pelad0}) we see that in the limit, all $K_0$'s vanish, all $\hat{K}_1$'s have finite limit, while $\mathcal{J}$ blows up logarithmically. Therefore, with logarithmic precision, we have
\begin{align}\label{pelad1}
\big \langle T_\nu^\nu\big\rangle_{\rm
ren}=\frac{1}{4 \pi^2 r^4}  A_{\nu,-1}(r=0)  \Big[\ln \frac{L}{r}+\mathcal{O}(1)\Big]\kn,
\end{align}
where coefficients $ A_{\nu,-1}(r)$ are given in the Table\,\ref{tabcoef} and one uses expansion (\ref{blatt4uu}) for $L=l_c$, and expansion (\ref{blatt4cf123}) for $L=d_s$.

 \begin{figure}
 \begin{center}
\includegraphics[width=10cm]{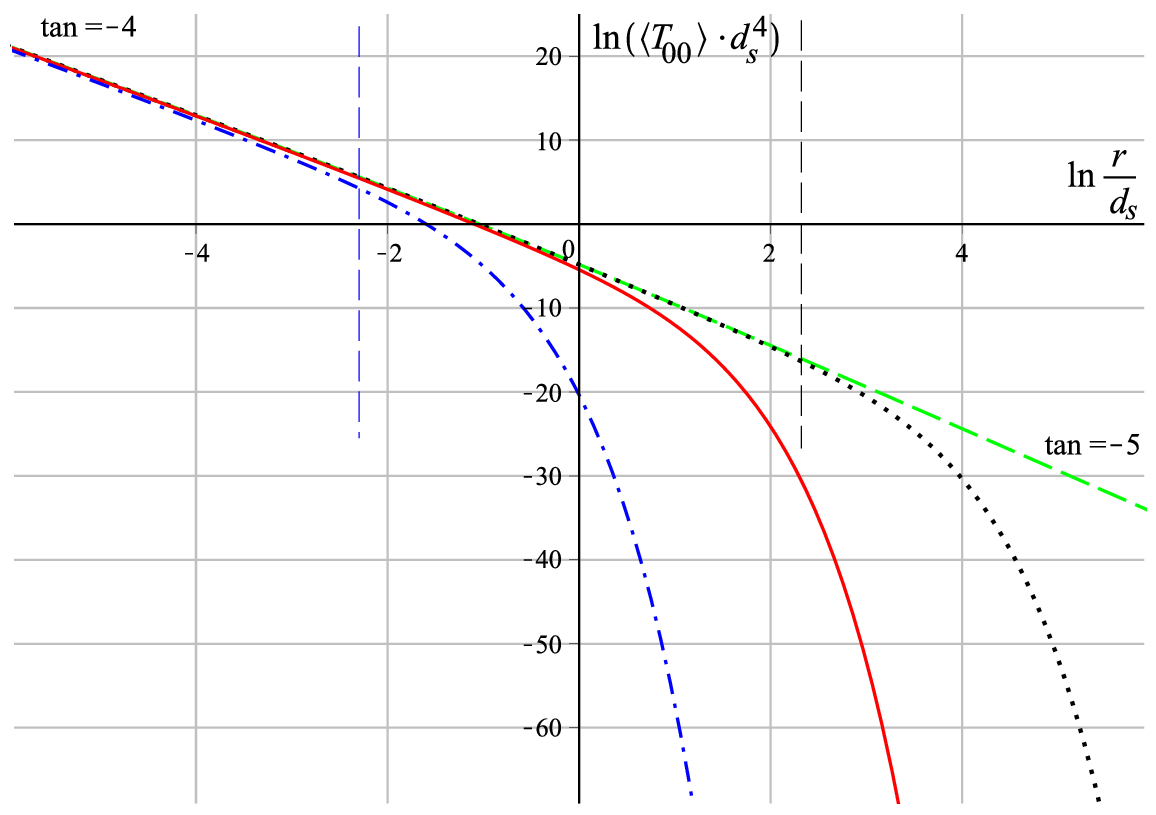}
 \caption{Renormalized vacuum density at small distances in doubly logarithmic scale (minimal coupling): for massless field (green dashed), for $l_c/d_s=10$ (black dotted), $l_c/d_s=1$ (red solid) and $l_c/d_s=0.1$ (blue dashdotted).}
 \label{T00log}
\end{center}
\end{figure}

The result is
\begin{align}\label{pelad2}
 \big\langle T_{ \mu}^{\nu}\big\rangle _{\rm ren}=
 \frac{1}{8 \pi^2 r^4}\,\ln \frac{L}{r}
\,\mathrm{diag}\Big(1-4\xi, 8 \xi-1 ,1-8\xi,1-8\xi\Big)\kn.
\end{align}

For the particular cases $\xi=1/4$ or $\xi=1/8$, the leading term will be given by a constant instead of a logarithm. There is no difficulty to carry out an expansion up to desirable order, based on the consistent asymptotic expansions given in the Appendix\,\ref{asymint}.

\medskip

\textit{To conclude}: basically, the components of renormalized EMT reproduce the  behavior of $\phisq$, up to two extra powers of $r$ in the denominator. The basic difference is an appearance of  the curvature coupling $\xi$ in the EMT.  Here we assume that the curvature coupling $\xi$  does not take large values. At the other hand, for some particular cases of $\xi$, the basic asymptotic may be changed for the specified components of EMT.

The plots of renormalized vacuum energy density $\langle \kn T_{0}^{0}\rangle _{\rm ren}$ (for minimal coupling, at scales   $\sim d_s$) are presented on the Fig.\,\ref{T00log} in doubly logarithmic scaling. It illustrates our deductions.

As $r \to 0$, all curves have limiting value $-4$ of tangent, in accord with (\ref{pelad2}). At large distances the massless curve has limiting tangent $-5$
as predicted by (\ref{Ttt5}), while all massive modes are exponentially suppressed at $r>l_c$. The points, corresponding to values $r=l_c$  for each massive curve, are marked by vertical dashed line of corresponding color. For all curves the value corresponding to $r=d_s$, is given by the cross with the ordinate axis.

\begin{figure}
 \begin{center}
\includegraphics[width=10cm]{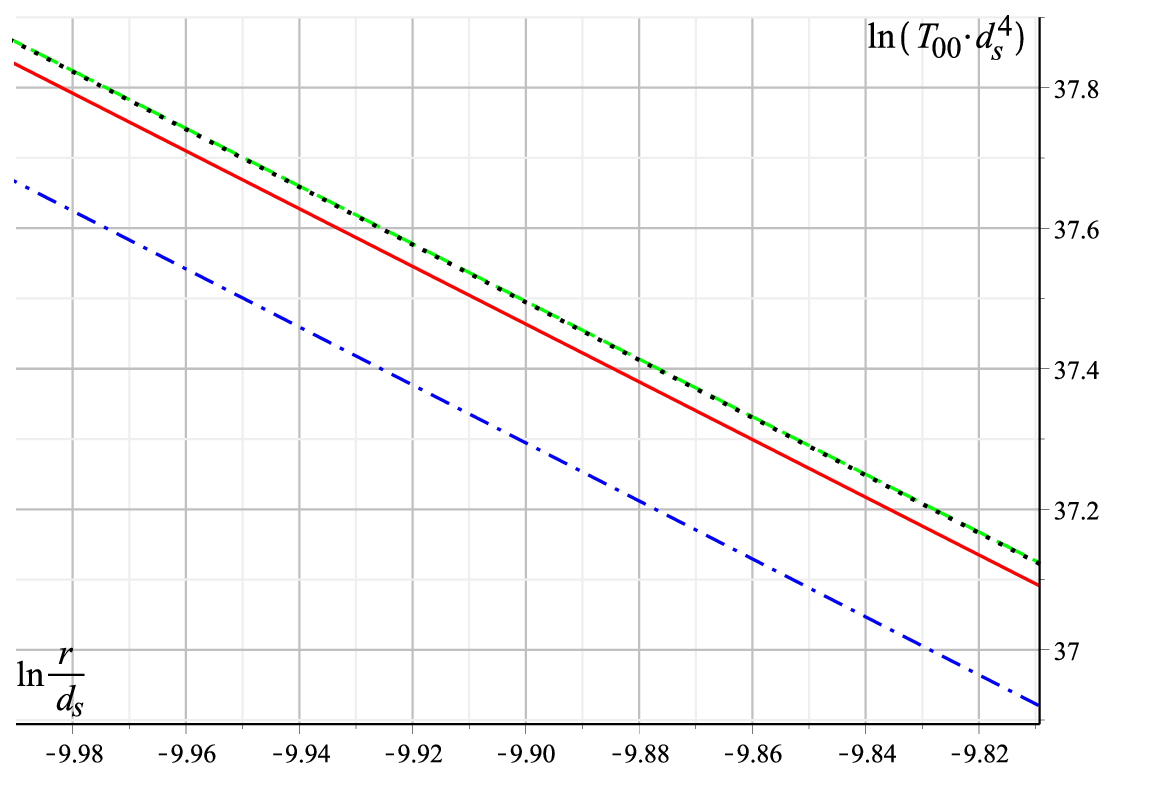}
 \caption{Renormalized vacuum density at small distances in doubly logarithmic scale (minimal coupling): for massless field (green dashed), for $l_c/d_s=10$ (black dotted), $l_c/d_s=1$ (red solid) and $l_c/d_s=0.1$ (blue dashdotted).}
 \label{T00}
\end{center}
\end{figure}

We see that at small ($r\lesssim l_c$) distances all curves, that correspond to massive fields (particles), are almost identical to the massless curve.
 The difference  is to be visible on the Fig.\,\ref{T00}, there we plot the same curves at very small $r\ll l_c$, $r\ll d_s$ distances.

All curves are almost parallel to each other, with gaps related with the logarithms of $l_c/r$ and $d_s/r$ (see eq.\,(\ref{pelad1})).
Therefore, the partial contributions of all modes (distinguished with respect to their mass) into the vacuum energy density, become comparable between themselves.

It confirms our conclusions made in the works \cite{Grats2023b,Grats2023c,Grats2024}, on the effectively-massless behavior of massive fields (in application to the Casimir effect) at small (with respect to the Compton length) distances.

\section{Conclusion}\label{Concl}

The problem of vacuum polarization of massive scalar field in the background of zero-range potential has been considered. Formally, the problem is reduced to the
consideration of three-dimensional stationary Schr\"{o}dinger equation with
$\delta$-like potential. However, the question how to regard the heuristically introduced $\lambda\kn \delta^3$-potential, is not trivial.
One might regard it as a limit of appropriate  $\delta$-like sequence, and work within the Perturbation-theory framework.  However, as it was shown by Berezin and Faddeev \cite{Berezin}, within this approach  the ''delta-like-strength'' coupling $\lambda$ has to be considered as bare and requires a renormalization. 
 In the work we made usage of an alternative technique, based on the  self-adjoint extensions for the operator
 \mbox{$-\Delta$}. For the repulsion potential, it allows to get exact expressions for the renormalized vacuum expectation values for $\phi^{\kn 2}$ and $T_{\mu\nu}$.
For the renormalization of the Green's function we apply subtraction scheme, and the sum over all modes degenerates in a single term,  due to the deformation of $s$-wave only. 

 The final result is determined by the self-adjoint-extension parameter, i.e. by the scattering length which uniquely characterizes a zero-range potential and may be observable  experimentally. All the renormalized quantities under interest are expressed in terms of a single two-parameter integral, which is studied in detail.

For three lengthy parameters in the problem~--- distance $r$, the Compton length $l_c$ and the scattering length $d_s$~--- we analyze a dependence of vacuum polarization effect with respect to each one. It is shown that at distances essentially smaller than $l_c$, the vacuum energy density of massive field becomes comparable with that one of the massless.
 At large distances the  vacuum energy density of massive field  decays exponentially, while for the massless field it falls as power-like function, and for massless conformal field the decay is faster.  The limit of the massless scalar field reproduces known non-perturbative results \cite{Grats2018}, and in particular, got by another regularization techniques \cite{Fermi1,Fermi2}. It justifies our choice of the Laplacian's self-adjoint extension as a proper description of the potential by pointlike impurity.

\medskip

\textbf{Acknowledgment.} This study was conducted within the scientific program of the National Center for Physics and Mathematics, section \#5 ''Particle Physics and Cosmology''. Stage 2023-2025.

\appendix
\section{The basic integral}\label{basint}
The basic integral under investigation is defined as
\begin{align}\label{origi}
\mathcal{J}(\beta,a ):=\int\limits_{0}^{\infty} d z\,\frac{1}{1+z^2}\,\frac{z}{\sqrt{z^2+a ^2}}
\Bigl(\sin\beta z+z\cos\beta z\Bigr)\, .
\end{align}
Introducing two additional integrals
\begin{align}
\mathcal{J}_{c}(\beta,a ):=\int\limits_{0}^{\infty} d z\,\frac{\cos\beta z}{1+z^2}\,\frac{1}{\sqrt{z^2+a ^2}}\, , \qquad\qquad
\mathcal{J}_{s}(\beta,a ):= \int\limits_{0}^{\infty} d z\,\frac{\sin\beta z}{1+z^2}\,\frac{z}{\sqrt{z^2+a ^2}}\, \nn
\end{align}
with mutual relation
\begin{align}
\mathcal{J}_{s}(\beta,a ) = -\frac{\partial \mathcal{J}_{c}(\beta,a )}{\partial \beta}\,,
 \nn
\end{align}
 the initial integral in term of the sine- and cosine-integrals takes the form
\begin{align}\label{blatt1}
\mathcal{J} = K_0(\beta a )+\mathcal{J}_{s} -\mathcal{J}_{c}\,,
\end{align}
where one  uses the  table integral
\begin{align}\label{tab1}
 \int\limits_{0}^{\infty} d z\, \frac{\cos\beta z}{\sqrt{z^2+a ^2}} = K_0(\beta a ) \,.
\end{align}

\begin{figure}
 \begin{center}
\includegraphics[width=7cm]{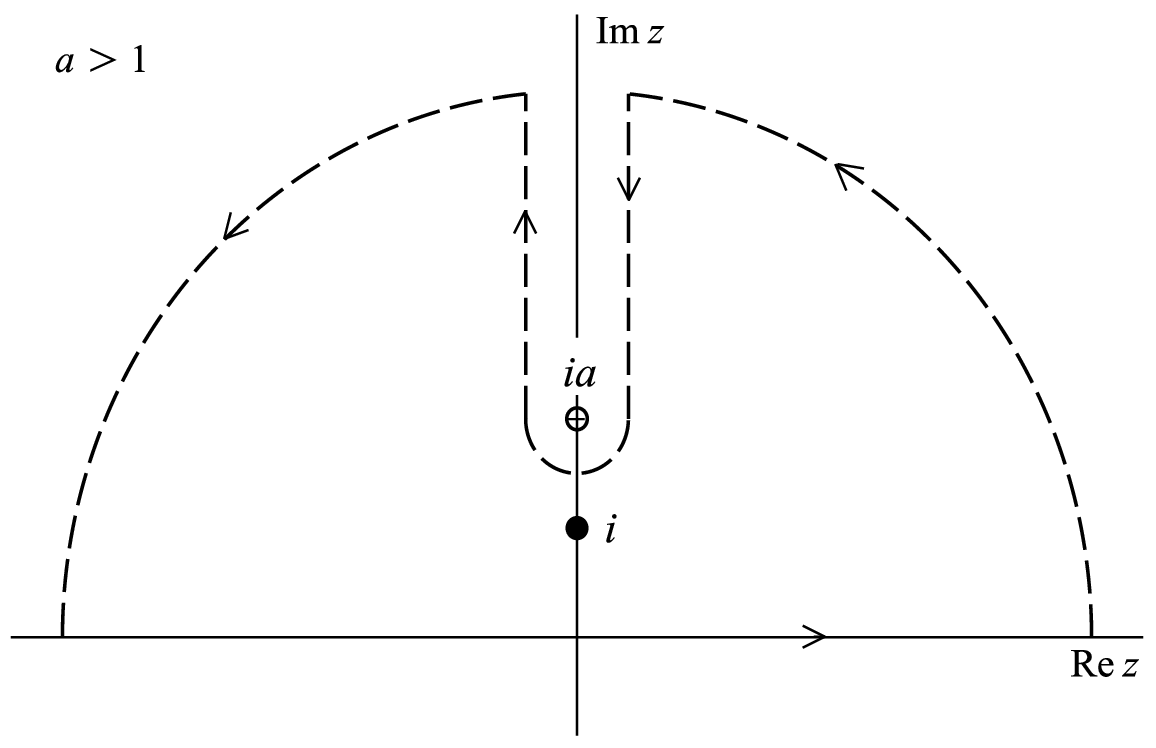}
 \caption{Integration contour.}
 \label{picx4a}
\end{center}
\end{figure}

Consider the  cos-integral $\mathcal{J}_{c}$. Expanding the integration domain to $ \mathbb{R}$ by parity, we represent it in the form
\begin{align}
\mathcal{J}_{c}(\beta,a )=\frac{1}{2}\int\limits_{0}^{\infty} d z\,\frac{\e^{i\beta z}}{1+z^2}\,\frac{1}{\sqrt{z^2+a ^2}}\, , \nn
\end{align}
with $\beta\geqslant 0$. Closing the integration contour in the upper half-plane of the complex variable  $z$, we find that the integrand has a simple pole at $z=i$ and a branching  point at $z=ia$.  The pole contribution is readily computed, while in order to avoid the branching point we make a cut from $ia$ to $i \infty$ (Fig.\,\ref{picx4a}). One can see that the pole part yields the real-valued contribution for $a>1$, and the cut-integral is finite; otherwise the cut goes near the pole and the cut-integral diverges.

Therefore the value
$a=1$ naturally splits the possible characteristic values of variable~$a$. Hence consider the cases   $a>1$, $a=1$ and $0\leqslant a<1$
separately.

\medskip

\il $\boldsymbol{a}\boldsymbol{>}\boldsymbol{1}.$
Implementing the scheme above, from the contour integral
 (Fig.\,\ref{picx4a}) we get:
\begin{align}
\mathcal{J}_{c}(\beta,a )= \frac{\pi}{2}\frac{\e^{-\beta}}{\sqrt{a^2-1}}- \int\limits_{a}^{\infty} d z\,\frac{\e^{-\beta z}}{z^2-1}\,\frac{1}{\sqrt{z^2-a ^2}}\,.\nn
\end{align}
Splitting $(z^2-1)^{-1}$ onto simple fractions, we represent each  fraction $(z\pm 1)^{-1}$ in the exponential from:
\begin{align}
\mathcal{J}_{c}(\beta,a )= \frac{\pi}{2}\frac{\e^{-\beta}}{\sqrt{a^2-1}}
-\frac{1}{2}\int\limits_{a}^{\infty} d z\int\limits_{0}^{\infty}dt\,\frac{\e^{-\beta z}}{\sqrt{z^2-a ^2}}\Big(\e^{-(z-1)t} - \e^{-(z+1)t}\Big)\,.\nn
\end{align}
Changing the integration sequence, the internal integral over $z$ reduces to the table integral (\ref{tab1}), transformed to the contour integral
along the same line as in Fig.\,\ref{picx4a}:
\begin{align}
 \int\limits_{a}^{\infty} d z \frac{\e^{- \sigma z}}{\sqrt{z^2-a ^2}} =K_0(\sigma a)\,.\nn
\end{align}
Then one obtains
\begin{align}\label{cos1}
\mathcal{J}_{c}(\beta,a )= \frac{\pi}{2}\frac{\e^{-\beta}}{\sqrt{a^2-1}}
- \int\limits_{0}^{\infty}dt\, \mathrm{sh}\,t\,K_0\big(a(t+\beta)\big)\,.
\end{align}
The latter expression still represents the single-variable integral (like an initial one), but the convergence is much more faster. Differentiating  (\ref{cos1}) with respect to $\beta$, the sine integral reads:
\begin{align}
\mathcal{J}_{s}(\beta,a ) = -\frac{\partial \mathcal{J}_{c}(\beta,a )}{\partial \beta}= \frac{\pi}{2}\frac{\e^{-\beta}}{\sqrt{a^2-1}}
- a \int\limits_{0}^{\infty}dt\, \mathrm{sh}\,t\,K_1\big(a(t+\beta)\big)\,.\nn
\end{align}
Integrating it by parts, one obtains
\begin{align}\label{sin1}
\mathcal{J}_{s}(\beta,a ) = \frac{\pi}{2}\frac{\e^{-\beta}}{\sqrt{a^2-1}}
-   \int\limits_{0}^{\infty}dt\, \mathrm{ch}\,t\,K_0\big(a(t+\beta)\big)\,.
\end{align}
Plugging (\ref{cos1}) and (\ref{sin1}) into eq.\,(\ref{blatt1}), we get
\begin{align}\label{blatt2}
\mathcal{J} = K_0(\beta a )-  \int\limits_{0}^{\infty}dt\, \e^{-t}\,K_0\big(a(t+\beta)\big)\,.
\end{align}
After the integration variable shift $t+b \longrightarrow t$, the integral under consideration reads eventually:
\begin{align}\label{blatt3}
\mathcal{J} = K_0(\beta a )- \e^{\kn \beta} \!\!\int\limits_{ \beta}^{\infty}dt\, \e^{-t}\,K_0 (at )\,.
\end{align}

\il  $\boldsymbol{a}\boldsymbol{=}\boldsymbol{1}.$ In this particular case both integrals are the table ones: the cosine one equals
\begin{align}
\mathcal{J}_{c}(\beta,1 )=\int\limits_{0}^{\infty} d z\,\frac{\cos\beta z}{(1+z^2)^{3/2}} = \beta K_1(\beta)\,.\nn
\end{align}
Upon differentiation, one obtains
\begin{align}
\mathcal{J}_{s}(\beta,1 )=\int\limits_{0}^{\infty} d z\,\frac{z\sin\beta z}{(1+z^2)^{3/2}} = \beta K_0(\beta)\,.\nn
\end{align}

Therefore,
\begin{align}\label{blatt4}
\mathcal{J}(\beta,1) = (1+\beta )K_0(\beta)  -\beta K_1(\beta)\,.
\end{align}

Exact value allows to reveal asymptotic behavior immediately.
For $ \beta\gg1 $ we use asymptotic expansion of Macdonald functions:
\begin{align}\label{macduck1}
K_{\nu}(x) =\sqrt{\frac{\pi}{2x}}\,\e^{-x}\sum_{k=0}^{\infty}\frac{\Gamma(\nu+k+1/2)}{\Gamma(\nu-k+1/2)}\,\frac{(2x)^{-k}}{k!}\,,
\end{align}
that yields
\begin{align}\label{macduck2}
 \mathcal{J}(\beta,1) =\sqrt{\frac{\pi}{8 \beta}}\,\e^{-\beta}\Big[1+\frac{1}{8\beta} -\frac{27}{128 \beta^2}+\mathcal{O}(\beta^{-3}) \Big]\kn.
\end{align}
Notice, the basic term of expansion (\ref{macduck1}) is identical for all indices $\nu$.

For small arguments $ \beta \ll 1 $ we use the following series of Macdonald functions:
\begin{align}\label{macduck3}
K_{0}(x) = \sum_{k=0}^{\infty}\frac{1}{(k!)^2} \Big(\frac{z}{2}\Big)^{\!2k}  \Big[  \ln\frac{2}{x}+h_k-\gamma \Big]\kn,
\end{align}
where $h_k$ stands for the $k$-th harmonic number, while $\gamma$ is the Euler\,--\,Mascheroni constant. The asymptotic for $K_{1}(x)$ is to be obtained by direct differentiation of that one for $K_{0}$, resulting in
\begin{align}\label{macduck4}
 \mathcal{J}(\beta,1) = \ln\frac{2}{\beta}-(\gamma+1) +   \Big[\ln\frac{2}{\beta}-\gamma  \Big] \beta +  \bigg[\frac{3}{4}\Big(\ln\frac{2}{\beta}-\gamma\Big)+\frac{1}{2}  \bigg] \beta^2+\mathcal{O}(\beta^3|\ln \beta|)\,.
\end{align}

\medskip

\il $\boldsymbol{a}\boldsymbol{<}\boldsymbol{1}.$
As it was mentioned above, the contour integration (Fig.\,\ref{picx4a}) yields diverging cut-integrals both for sine-and for cosine integrals. Whereas the cut
from
0 to $ia$ allows to compute only the odd-integrand expressions. Meanwhile, we need sine- and   cosine integrals as their difference only. Taking look on  (\ref{blatt3}), we find that it converges at any value
of parameter $a$. Therefore, for any $a>0$ we can use representation
\begin{align}\label{blatt3vv}
\mathcal{J}(\beta,a) = K_0(\beta a )- \e^{\kn \beta} \!\!\int\limits_{ \beta}^{\infty}dt\, \e^{-t}\,K_0 (at )\,.
\end{align}
In contrast with the initial expression (\ref{origi}), it (i) allows to directly expand in large and small parameters  $a$ and $\beta$ with no appellation to the matched asymptotic expansion; (ii) allows to differentiate it twice, to compute the Energy-Momentum tensor; (iii) has rapidly decaying integrand, what may be used e.g. in numerical integration.

\medskip

 \textbf{Derivatives.} For the computation of renormalized Energy-Momentum tensor we need derivatives with respect to argument $\beta$. For convenience, introduce the following functions:
 \begin{align}
\mathcal{J}_1(\beta,a):=\beta \frac{\partial \mathcal{J}(\beta,a)}{\partial \beta}\,, \qquad\qquad \mathcal{J}_2(\beta,a):=\beta^2 \frac{\partial^2\! \mathcal{J}(\beta,a)}{\partial \beta^2}\,.\nn
\end{align}
Upon direct differentiation, one can deduce:
\begin{align}
\mathcal{J}_1(\beta,a)=\beta \mathcal{J}  - \hat K_1(\beta a) \,
,\qquad\qquad
\mathcal{J}_2(\beta,a)= \beta^2\mathcal{J} -(\beta  -1) \hat K_1(\beta a)+(\beta a)^2 K_0(\beta a )\,.\nn
\end{align}
Here $ \hat K_{\nu}(x):=   x^{\nu} K_{\nu}(x)$ (non-conventional but useful notation!) has properties: $$\hat K_{\nu}(0)=2^{\nu-1}\Gamma(\nu)\,, \qquad\qquad \big[ \hat {K}_{\nu}(x)\big]'=-x  \hat{K}_{\nu-1}(x)\,. $$

As $a\to 0^+$, the limits of $\mathcal{J}_k$ transit to the corresponding functions, considered in \cite{Graz21}:
\begin{align}\label{limsa}
\mathcal{J}(\beta,0)=\e^{\beta} E_1(\beta) \,
,\qquad\quad\mathcal{J}_1(\beta,0)=\beta \mathcal{J}(\beta,0)  - 1 \,
,\qquad\quad
\mathcal{J}_2(\beta,0)= \beta^2\mathcal{J}(\beta,0)-\beta  +1 \,,
\end{align}
where
\begin{align}
E_1(x):=\int\limits_x^{\infty}\frac{\e^{-t}}{t}\,dt\nn
\end{align}
denotes the integral exponent.

\section{Asymptotic behavior of the basic integral}\label{asymint}

Upon integration by parts in the expression (\ref{blatt3vv}), one may reduce the basic integral $\mathcal{J}(\beta,a)$ to the more economy form
\begin{align}\label{blatt-alter}
\mathcal{J}(\beta,a) = a \kn\e^{\kn \beta} \!\!\int\limits_{ \beta}^{\infty}dt\, \e^{-t}\,K_1 (at )\,.
\end{align}

If convenient, $\mathcal{J}(\beta,a)$ in (\ref{blatt3vv}) may be converted into the integral over domain $[0;\beta]$, using the table integrals~\cite{Ryzhik}:
\begin{align}
\int\limits_{0}^{\infty}dt\, \e^{-t}\,K_0 (at )=  \left\{
                                                    \begin{array}{ll}
                                                      \mathrm{Arch}\,a^{-1}/\sqrt{1-a^2}, & \hbox{$a<1$;} \\
                                                      1, & \hbox{$a=1$;} \\
                                                      \mathrm{arccos}\,a^{-1}/\sqrt{a^2-1}, & \hbox{$a>1$.}
                                                    \end{array}
                                                  \right.\nn
\end{align}
Then for   $ \beta<1$ the following representation is useful:
\begin{align}\label{blatt4fff}
\mathcal{J}(\beta,a) = K_0(\beta a )+ \e^{\kn \beta} \!\!\int\limits_0^{ \beta} dt\, \e^{-t}\,K_0 (at )- \frac{ \e^{\kn \beta}}{\sqrt{|1-a^2|}} \,\bigg\{
\acs=1pt\begin{array}{l}
  \mathrm{Arch}  \\
 \mathrm{arccos}
\end{array}
\bigg\}\frac{1}{a}\,,
\end{align}
where the inverse cosine function is to be chosen according to the expression's real value. It will be useful for expansions with small $a$.

 The primary factor that determines an asymptotic behavior, is the
 product $a\beta$
inside the Macdonald functions, hence consider the limiting cases of this product.

\medskip

\mbox{\textbf{I. Case $\boldsymbol{a}\boldsymbol{\beta}\boldsymbol{\ll}\mathbf{1}$.}} In this case the argument of both Macdonald functions in the representation (\ref{blatt4fff}) is much smaller than unity, hence one uses the small-argument asymptotic (\ref{macduck3})  for $K_0$:
\begin{align}\label{blatt4cf}
\mathcal{J}(\beta,a) = \ln \frac{2}{a \beta} -\gamma +
\e^{\beta}\!\int\limits_0^{ \beta} dt\, \e^{-t}\Big[\ln \frac{2}{a t}  -\gamma \Big] - \frac{ \e^{\kn \beta}}{\sqrt{|1-a^2|}} \,\bigg\{
\acs=1pt\begin{array}{l}
  \mathrm{Arch}  \\
 \mathrm{arccos}
\end{array}
\bigg\}\frac{1}{a} +\mathcal{O}(a^2\beta^2 |\ln a \beta|) \,.
\end{align}

\il  For
$\beta\ll a \sim 1$ on the integration domain one has  $\e^{-t}\simeq 1$; thus we integrate (\ref{blatt4cf}) with respect to $t$ and expand the result with respect to small $\beta$:
\begin{align}\label{blatt4uu}
\mathcal{J}(\beta,a) = \ln \frac{2}{a \beta} -\gamma-\frac{ 1+\beta}{\sqrt{|1-a^2|}} \,\bigg\{
\acs=1pt\begin{array}{l}
  \mathrm{Arch}  \\
 \mathrm{arccos}
\end{array}
\bigg\}\frac{1}{a}+
\beta\Big[\ln \frac{2}{a \beta} +1-\gamma \Big] +\mathcal{O}(\beta^2 |\ln  \beta|) \,.
\end{align}

\il  For $a\ll \beta \sim 1$  after the expansion inside the integrand (\ref{blatt4cf})
one encounters the following family of integrals:
\begin{align}
\Lambda_n(\beta):= \int\limits_0^{ \beta} \e^{-t}  t^n \,\ln t \,dt\,.\nn
\end{align}
Introducing the extra-parameter family
\begin{align}
\tilde{\Lambda}_n(\beta,c):= \int\limits_0^{ \beta} \e^{-ct}  t^n \,\ln t \,dt\,, \qquad\qquad \Lambda_n(\beta)=\tilde{\Lambda}_n(\beta,1)\,,\nn
\end{align}
with recurrence
\begin{align}
\tilde{\Lambda}_n(\beta,c):= -\frac{\partial}{\partial c}\tilde{\Lambda}_{n-1}(\beta,c)\,,\nn
\end{align}
hence we need just the generating one $\tilde{\Lambda}_0(\beta,c)$:
\begin{align}
\tilde{\Lambda}_0(\beta,c) =\int\limits_0^{ \beta} \e^{-ct}  \ln t \,dt=\bigg[\int\limits_0^{ \infty} -\int\limits_{\beta}^{ \infty} \bigg] \e^{-ct}  \ln t \,dt\,.\nn
\end{align}
The first integral is a table one, in the second the integration by parts reduces it to the form of $E_1$:
\begin{align}
\tilde{\Lambda}_0(\beta,c) = \frac{1}{c}\,\Big[E_1(\beta c)+\e^{-\beta c}\ln \beta+\gamma+\ln c \Big]\,.\nn
\end{align}
Implementing the scheme, the lowest necessary $\Lambda$'s are:
\begin{align}
&  \Lambda_0(\beta)=-\big[E_1(\beta)+\gamma+\e^{-\beta}\ln \beta\big]\,, \nn \\
&  \Lambda_2(\beta)= 3(1-\e^{-\beta})-2\Big[E_1(\beta)+\gamma -\e^{-\beta}\ln \beta(\beta+1)\Big]-\beta\e^{-\beta}\big(1+\beta\ln \beta\big)\kn.\nn
\end{align}
Therefore, the expansion is given by
\begin{align}\label{blatt4dd}
\mathcal{J}(\beta,a) =\e^{\kn \beta} E_1(\beta) -  \frac{\beta+1}{2}\,a^2 \ln\frac{2}{a}+
 \Big[2\e^{\kn \beta} E_1(\beta) +(2\gamma-1)(\beta-1)+(\beta^2+2\beta+2)\ln \beta\Big]    \frac{a^2}{4}   +    \mathcal{O}(a^4 |\ln a|) \,.
\end{align}

\il  Finally, for $a\ll 1$, $ \beta\ll 1$, one makes use of the asymptotic expansion
 \cite{Ryzhik}
\begin{align} \label{E1small}
E_1(x)=|\ln x|-\gamma -
\sum\limits_{k=1}^{\infty}\,\frac{(-x)^k}{k\cdot k!}\,,\qquad\qquad 0<x\ll 1\,,
\end{align}
to get the double expansion:
\begin{align}\label{blatt7}
\mathcal{J}(\beta,a) =\Big(\ln \frac{1}{\beta}-\gamma\Big)(1+\beta) +\beta   -\frac{1}{2} \Big(\beta^2\ln\frac{1}{\beta}+a^2\ln\frac{1}{a}\Big)  + \mathcal{O}(a^2 +\beta^2) \,.
\end{align}
Thus if   $a$ and $\beta$ are of the same order, then contribution due to  $\beta$ dominates.

\il For $a \ll 1$, $ \beta \gg 1$, $ a\beta \ll 1$  one may take advantage of all the transformations leading to eq.\,(\ref{blatt4dd}). Then for  $ \beta \gg 1$ one uses the asymptotic series of the integral exponent\footnote{The series diverges but being truncated at $n=  [ x]$, it yields the exponentially small error.}
\begin{align}\label{Ei_asy}
\e^x E_1(x)=\frac{1}{x}\sum\limits_{k=0}^{\infty}\frac{(-1)^{k}\,k!}{x^k}\,, \qquad\qquad x\gg1
\end{align}
to obtain
\begin{align}\label{blatt6a}
\mathcal{J}(\beta,a) =\frac{1}{\beta}- \frac{1}{\beta^2} +    \mathcal{O}\big( \beta^{-3}+a \beta^{-1} |\ln a|\big) \,.
\end{align}
This case is qualitatively similar to $a=0$, given by expression (\ref{limsa}).

\il For $a \gg1$, $ \beta \ll 1$, $ a\beta \ll 1$ we start with  (\ref{blatt4uu}) and just expand  arccos with respect to $a^{-1}$:
\begin{align}\label{blatt4cf123}
\mathcal{J}(\beta,a) = \ln \frac{2}{a \beta} -\gamma- \frac{ \pi}{2a} +
\beta\Big[\ln \frac{2}{a \beta} +1-\gamma \Big] +\mathcal{O}(a^2\beta^2 |\ln  a\beta|) \,.
\end{align}

\medskip

\textbf{II. Case $\boldsymbol{a}\boldsymbol{\beta}\boldsymbol{\gg}\mathbf{1}$.} If $a\beta\gg 1$, then we use representation  (\ref{blatt3vv}) and asymptotic  (\ref{macduck1}):
\begin{align}
\mathcal{J} = K_0(\beta a )-\sqrt{\pi} \,\e^{\kn \beta} \!\!\int\limits_{ \beta}^{\infty}dt\, \e^{-(a+1)t} \sum_{k=0}^{\infty}\frac{\Gamma(k+1/2)}{k!\,\Gamma(-k+1/2)}\,\frac{1}{(2at)^{k+1/2}}\,.\nn
\end{align}
Changing the summation-integration sequence  and introducing the integrals
\begin{align}
\Phi_k(\sigma,\beta):=\int\limits_{ \beta}^{\infty}dt\, \frac{\e^{- \sigma t}}{t^{k+1/2}} \,,\nn
\end{align}
we get
\begin{align}\label{blatt10}
\mathcal{J} = K_0(\beta a )-\sqrt{\pi} \,\e^{\kn \beta}  \sum_{k=0}^{\infty}\frac{\Gamma(k+1/2)}{k!\,\Gamma(-k+1/2)}\,\frac{\Phi_k(a+1,\beta)}{(2a)^{k+1/2}}\,.
\end{align}
Let study $\Phi_k(\sigma,\beta)$: for $k=0$ the result is given by table integral:
\begin{align}
\Phi_k(\sigma,\beta)=\sqrt{\frac{\pi}{\sigma}}\,\mathrm{erfc}(\sqrt{\sigma\beta})\,, \qquad\qquad \mathrm{erfc}(\cdot):=1-\mathrm{erf}(\cdot)\,,\nn
\end{align}
where ''$\mathrm{erf}$'' stands for the error function, while for $k\geqslant 1$ one integrates by parts to obtain the recurrence:
\begin{align}
\Phi_k(\sigma,\beta)=\frac{\e^{-\sigma\beta}}{(k-1/2)\beta^{k-1/2}}-\frac{\sigma}{k-1/2}\,\Phi_{k-1}(\sigma,\beta)\,.\nn
\end{align}
After $k$-th integration by parts we have
\begin{align}\label{blatt13}
\Phi_k(\sigma,\beta)&=\frac{\e^{-\sigma\beta}}{\beta^{k-1/2}} \sum_{l=0}^{k-1}(-1)^l \frac{\Gamma(k-l-1/2)}{\Gamma(k+1/2)} \,(\sigma\beta)^l+
\frac{(-1)^k \sigma^{k}}{\Gamma(k+1/2)}\,\sqrt{\pi}  \,\Phi_{0}(\sigma,\beta)=\nn\\
&=\frac{\e^{-\sigma\beta}}{\beta^{k-1/2}} \sum_{l=0}^{k-1}(-1)^l \frac{\Gamma(k-l-1/2)}{\Gamma(k+1/2)} \,(\sigma\beta)^l+
\frac{(-1)^k \sigma^{k-1/2}}{\Gamma(k+1/2)}\, \pi   \, \mathrm{erfc}(\sqrt{\sigma\beta})\,.
\end{align}
Now it is clear that the product $\beta^{k-1/2}\Phi_k(\sigma,\beta)$ depends upon the product $\sigma\beta\gg1$ only. Plugging the asymptotic expansion of the
''complement error function '' ($\mathrm{erfc}$)
\begin{align}\label{blatt14}
\mathrm{erfc}(z)=\frac{\e^{-z^2}}{\pi}\sum_{m=0}^{\infty}\frac{(-1)^m\Gamma(m+1/2)}{z^{2m+1}}\,,\qquad\qquad z\gg1
\end{align}
and changing the summation index  as  $l=k-1-m $ in the sum (\ref{blatt13}), we find that the first   $k$ terms of the expansion (\ref{blatt14}) precisely cancel all the terms of sum (\ref{blatt13}). Therefore,
\begin{align}
\Phi_k(\sigma,\beta)=
\frac{(-1)^k \sigma^{k-1/2}}{\Gamma(k+1/2)}\,   \e^{- \sigma\beta} \sum_{m=k}^{\infty}\frac{(-1)^m\Gamma(m+1/2)}{(\sigma\beta)^{m+1/2}}\,.\nn
\end{align}
In what follows,   for all $k\geqslant 0$ we have
\begin{align}
\Phi_k(\sigma,\beta)=
   \frac{\e^{- \sigma\beta}}{\sigma \beta^{k+1/2}}  \Big[ 1-\frac{k+1/2}{\sigma\beta} +\mathcal{O}\left((\sigma\beta)^{-2}\right) \Big]\,.\nn
\end{align}
Substituting it into eq.\,(\ref{blatt10}), the leading terms may be summed up back with help of (\ref{macduck1}), to yield
\begin{align}\label{blatt17}
 \mathcal{J}   \simeq\frac{a}{a+1}\, K_0(a\beta   )  \,.
\end{align}
For consistent series, we have to expand Macdonald function and take into account all corrections:
\begin{align}\label{blatt17corr}
 \mathcal{J}  =   \sqrt{\frac{\pi}{2a\beta }}\,\e^{-a\beta }\bigg[ \frac{a}{a+1} +\frac{3-a}{8(a+1)^2 \beta}+      \mathcal{O}\left((a\beta)^{-2}\right) \bigg]\kn.
\end{align}

The latter is also applicable to the case $\beta\gg1$, $a\sim 1$.

\il In particular, if $a\gg1$, then
\begin{align}\label{blatt18}
 \mathcal{J}     \simeq   \sqrt{\frac{\pi }{2a\beta }}\,\e^{-a\beta }\Big[ 1+\mathcal{O}\left((a\beta)^{-1}\right) \Big]\kn.
\end{align}
This variant covers the cases $\beta\sim 1$, $\beta\gg 1$ and $a\gg 1/\beta\gg1$.

\il For $a \ll 1$, $ \beta \gg 1$, $ a\beta \gg 1$ one uses  all arguments, which lead to expression (\ref{blatt17corr}). Expanding it with respect to small $a$, we get
\begin{align}\label{blatt17nn}
 \mathcal{J}  =  \sqrt{\frac{\pi}{2a\beta }}\,\e^{-a\beta }\bigg[ a+  \Big(\frac{3}{8\beta}-a^2\Big)+  \Big(a^3-\frac{7}{8}\frac{a}{\beta}-\frac{15}{128}\frac{1}{ a\beta^2}\Big) +\pp   \bigg]\kn,
\end{align}
where round parentheses denote the relative expansion order for the case $\alpha^2 \beta\sim 1$.

\medskip

\textbf{III. Case $\boldsymbol{a}\boldsymbol{\beta}\boldsymbol{\sim}\mathbf{1}$.} Finally, consider also the transition cases: where $a\gg 1 \gg \beta$ and \textit{vice versa}.

\medskip

\il For $a \ll1$, $ \beta \gg 1$, $ a\beta \sim 1$ we use the representation (\ref{blatt2}). Exponential decay of the whole integrand is determined by the arguments both of the exponent and of the Macdonald function. The latter factor reveals the domain   $t\lesssim 1/a \gg1$.

Therefore, in a domain which contributes basically, the Macdonald function varies very slowly, with respect to the change of  argument $t$. Expand  the Macdonald function in the neighbourhood of $a \beta$:
\begin{align}\label{blatt2jj}
\mathcal{J} = K_0(\beta a )-  \int\limits_{0}^{\infty}dt\, \e^{-t} \Big( K_0 (a\beta)  - a  t\kn K_1 (a\beta) +\frac{a^2  t^2}{4} \big[K_0 (a\beta)+K_2(a\beta)\big]   +\mathcal{O}\big(a^3\big) \Big)\kn.
\end{align}
In other words, in the valuable domain the Taylor expansion is valid, while the domain $t\gtrsim 1/a$, where the Taylor expansion works improperly, yields the exponentially suppressed contribution due to $\e^{-t}$.

Thus after the  $t$-integration one obtains  the consistent expansion in powers of small $a\ll1$. The leading term in this expansion cancels the terminal term (\ref{blatt2jj}), and therefore the leading surviving term is determined by the first correction:
\begin{align}\label{blatt2kk}
\mathcal{J} =  a \kn K_1 (a\beta)-\frac{a^2  }{2} \big[K_0 (a\beta)+K_2(a\beta)\big]+\frac{a^3  }{4} \big[3 K_1 (a\beta)+K_3(a\beta)\big] + \mathcal{O}(a^4) \,.
\end{align}

\il For  $a \gg1$, $ \beta \ll 1$, $ a\beta \sim 1$ it is more useful to work with integral term in the form  (\ref{blatt2}). Since $ a\beta \sim 1$, the exponential ''cut-off'' of Macdonald function takes place at $t\gtrsim 1/a$, hence the most valuable contribution is taken from the integration domain   $0<t\lesssim a^{-1}\ll 1$. On this area one effectively fixes $\e^{-t}\simeq 1+\mathcal{O}(t)$:
\begin{align}
\mathcal{J}  \simeq K_0(\beta a )-  \int\limits_{0}^{\infty}dt\, K_0\big(a(t+\beta)\big)\,.\nn
\end{align}
Shifting the integration variable as  $t \longrightarrow t-\beta$, transforming the integration domain to $[0,\beta]$ and rescaling the new integration variable as $t \longrightarrow \beta t$, we have:
\begin{align}
\mathcal{J} = K_0(a\beta ) - \frac{\pi}{2a}+ \beta\int\limits_0^{1} dt\, K_0 (a \beta t )+\mathcal{O}(\beta^2) \,.\nn
\end{align}
It may be integrated up to the modified Struve functions ($\text{\textsf{L}}_{\nu}$):
\begin{align}\label{blatt2yyy}
\mathcal{J} = (1+\beta)K_0(a\beta )+ \frac{\pi \beta}{2}\Big(\text{\textsf{L}}_0(a\beta ) K_1(a\beta ) +\text{\textsf{L}}_1(a\beta )K_0(a\beta )   -\frac{1}{a \beta}\Big)+\mathcal{O}(\beta^2) \,.
\end{align}

\medskip

The leading terms of asymptotic expansions obtained above, are combined in the Table\,\ref{logor}.

\begin{table}[h]
\begin{center}\small \extrarowheight=7pt \tcs=0mm
\begin{tabular}{c p{1cm} c>{\hspace*{4.5mm}}c>{\hspace*{6mm}}c>{\hspace*{6mm}}c>{\hspace*{4.5mm}}c}
\toprule
  \multirow{2}*{
  \begin{tabular}{l}
 \footnotesize Typical \\[-12pt]
   \footnotesize variants \\[-12pt]
  \footnotesize  of values of $ a$
  \end{tabular}
 } & & \multicolumn{5}{c}{\footnotesize Characteristic limiting values of $\beta$} \\\cline{3-7}
  & &\footnotesize $ \beta\akn\ll \akn a^{-1}\!\!\ll \akn 1$ & \footnotesize $ \beta\ll 1$ &\footnotesize $\beta\sim 1$ &  \footnotesize $\beta \gg1$ & \footnotesize  $\beta\!\gg\akn a^{-1}\!
\!\gg\akn 1$
  \\  \midrule
\addlinespace[1.1pt]
  $a=0$ & & --- & $\ds\ln \frac{1}{\beta} -\gamma $& $\ds \e^{\beta}E_1(\beta)$ & $1/\beta$ & --- \\[7pt]
\rowcolor{superlight-gray}
  $a\ll 1$ & & --- & $\ds \Big(\ln \frac{1}{\beta}-\gamma\Big)(1+\beta)+\beta$ & $ \ds \e^{\beta}E_1(\beta)\akn -\akn \frac{\beta\akn +\akn 1}{2}\,a^{\akn2}\ln\frac{2}{a}$  &  $a K_1(a\beta)$& $a K_0(a\beta)$ \\[7pt]
$a< 1$ & & --- & $ \ds\ln \frac{2}{a\beta}-\gamma-\frac{\mathrm{Arch}\,a^{-1}}{\sqrt{1-a^2}} $ & $\ds a \kn \e^{\beta}\!\! \int_{ \beta}^{\raisebox{2pt}{$\scriptstyle\infty$}}\!\!\!\!dt\, \e^{-t}K_1 (at )$  &  $\ds\sqrt{\frac{\pi a}{2\beta}}\frac{\e^{-a\beta}}{a+1}$& --- \\[7pt]
\rowcolor{superlight-gray}
$a= 1$ & & --- & $ \ds\ln \frac{2}{\beta}-\gamma-1 $ & $\ds(1+\beta)K_0(\beta)- \beta{K}_1(\beta)  $  &  $\ds\sqrt{\frac{\pi }{8\beta}}\, \e^{-\beta} $& --- \\[6pt]
$a> 1$ & & --- & $\ds\ln \frac{2}{a\beta}-\gamma-\frac{\mathrm{arccos}\,a^{-1}}{\sqrt{a^2-1}} $ & $\ds a\kn \e^{\beta}\!\! \int_{ \beta}^{\raisebox{2pt}{$\scriptstyle\infty$}}\!\!\!\!dt\, \e^{-t}K_1 (at )$  &  $\ds\sqrt{\frac{\pi a}{2\beta}}\frac{\e^{-a\beta}}{a+1} $& --- \\[7pt]
\rowcolor{superlight-gray}
$a\gg 1$ & &$\ds\ln \frac{2}{a\beta}\akn -\akn \gamma\akn -\frac{ \pi}{2a} $ & $K_0(a\beta) $ & $\ds\sqrt{\frac{\pi }{2a\beta}}\, \e^{-a\beta}\vphantom{\frac{A}{A_{\frac{q}{q}}}}  $  &  $\ds\sqrt{\frac{\pi }{2a\beta}}\, \e^{-a\beta}\vphantom{\frac{A}{A_{\frac{q}{q}}}}  $& --- \\
\addlinespace[-1.4pt]
  \bottomrule
\end{tabular}\caption{Asymptotic behavior  of the basic integral  $\mathcal{J}(\beta,a) $ with respect to both variables}\label{logor}
\end{center}
\end{table}

\end {document}